\documentclass[preprint,12pt,authoryear]{elsarticle}
\usepackage[utf8]{inputenc}
\usepackage{latexsym} 
\usepackage{graphicx}
\usepackage{color}
\usepackage{amssymb}
\usepackage{amsmath}
\usepackage{amsfonts}
\usepackage{overpic}
\usepackage{subcaption}

\newcommand{\pa}[1]{\left(#1\right)}

\newcommand{\newc}{\newcommand}
\newc{\beq}{\begin{equation}}
\newc{\eeq}{\end{equation}}
\newc{\kt}{\rangle}
\newc{\br}{\langle}
\newc{\beqa}{\begin{eqnarray}}
\newc{\eeqa}{\end{eqnarray}}
\newc{\pr}{\prime}
\newc{\longra}{\longrightarrow}
\newc{\ot}{\otimes}
\newc{\rarrow}{\rightarrow}
\newc{\h}{\hat}
\newc{\bom}{\boldmath}
\newc{\btd}{\bigtriangledown}
\newc{\al}{\alpha}
\newc{\be}{\beta}
\newc{\ld}{\lambda}
\newc{\sg}{\sigma}
\newc{\p}{\psi}
\newc{\eps}{\epsilon}
\newc{\om}{\omega}
\newc{\mb}{\mbox}
\newc{\tm}{\times}
\newc{\hu}{\hat{u}}
\newc{\hv}{\hat{v}}
\newc{\hk}{\hat{K}}
\newc{\ra}{\rightarrow}
\newc{\non}{\nonumber}
\newc{\ul}{\underline}
\newc{\hs}{\hspace}
\newc{\longla}{\longleftarrow}
\newc{\ts}{\textstyle}
\newc{\f}{\frac}
\newc{\df}{\dfrac}
\newc{\ovl}{\overline}
\newc{\bc}{\begin{center}}
\newc{\ec}{\end{center}}
\newc{\dg}{\dagger}
\newc{\prh}{\mbox{PR}_H}
\newc{\prq}{\mbox{PR}_q}
\newc{\tr}{\mbox{tr}}
\newc{\pd}{\partial}
\newc{\qv}{\vec{q}}
\newc{\pv}{\vec{p}}
\newc{\dqv}{\delta\vec{q}}
\newc{\dpv}{\delta\vec{p}}
\newc{\mbq}{\mathbf{q}}
\newc{\mbqp}{\mathbf{q'}}
\newc{\mbpp}{\mathbf{p'}}
\newc{\mbp}{\mathbf{p}}
\newc{\mbn}{\mathbf{\nabla}}
\newc{\dmbq}{\delta \mbq}
\newc{\dmbp}{\delta \mbp}
\newc{\T}{\mathsf{T}}
\newc{\J}{\mathsf{J}}
\newc{\sfL}{\mathsf{L}}
\newc{\C}{\mathsf{C}}
\newc{\B}{\mathsf{M}}
\newc{\V}{\mathsf{V}}
\makeatletter
\providecommand*{\diff}%
        {\@ifnextchar^{\DIfF}{\DIfF^{}}}
\def\DIfF^#1{%
        \mathop{\mathrm{\mathstrut d}}%
                \nolimits^{#1}\gobblespace
}
\def\gobblespace{%
        \futurelet\diffarg\opspace}
\def\opspace{%
        \let\DiffSpace\!%
        \ifx\diffarg(%
                \let\DiffSpace\relax
        \else
                \ifx\diffarg\[%
                        \let\DiffSpace\relax
                \else
                        \ifx\diffarg\{%
                                \let\DiffSpace\relax
                        \fi\fi\fi\DiffSpace}
\makeatother

\begin{document}

\begin{frontmatter}

\title{Records in the classical and quantum standard map}
\author[pks]{Shashi C. L. Srivastava\fnref{fn1}}
\address[pks]{Max-Planck-Institut f\"ur Physik komplexer Systeme,
N\"othnitzer Str.\ 38, 01187 Dresden, Germany}
\fntext[fn1]{On leave from Variable Energy Cyclotron
Centre, Kolkata, India.}
\author[iit]{Arul Lakshminarayan}
\address[iit]{Department of Physics, Indian Institute of Technology
Madras, Chennai 600036, India.}

\begin{abstract}
Record statistics is the study of how new highs or lows are created
and sustained in any dynamical process. The study of the highest or
lowest records constitute the study of extreme values. This paper
represents an exploration of record statistics for certain aspects of
the classical and
quantum standard map. For instance the momentum square or energy
records is shown to behave like that of records in random walks
when the classical standard map is in a regime of hard chaos. However
different power laws is observed for the mixed phase space
regimes. The presence of accelerator modes are well-known to create
anomalous diffusion and we notice here that the record statistics is
very sensitive to their presence. We also discuss records in random vectors and use it to analyze the {\it quantum} standard map via records in their eigenfunction intensities, reviewing some recent results along the way.
\end{abstract}

\end{frontmatter}


\section{Introduction}\label{reco:intro}

Breaking a record, or setting a new one has been a human passion, perhaps one
may say weakness, for a while now. Evidence how one yearns for a record to be
set by a particular sportsperson or the euphoria (depression) that sets in
when the stock market fluctuations reach a maximum (minimum) never seen
before.  Of course our view of nature also abounds in such characterizations:
the highest mountains, the deepest oceans, the smallest temperatures reached
and so on. While these may seem to be questions about extremes, records are a
record of extremes, the times they last and the number of new ones
created. For engineers designing a dam, one important piece of information is
when and by how much the water level of the river has exceeded all its
previous values; similar questions are naturally asked about rainfall when
planning for agricultural policies is taken up.  Practical considerations such
as these have been the early applications in a study of events which exceed
themselves in some aspect and are often called as ``record statistics", akin
to the study of extreme statistics in many ways. An accessible introduction to
record statistics is found in \cite{SchZia1999}.

 Questions about how the records increase with time, or the number of records
set, are of natural interest in all these complex, sometimes social contexts,
 have therefore been studied for example, \cite{gts02n,vkm01wrr,rp06pre}.
  A mathematical theory of records for
independent identically distributed ({\it i.i.d.}) random variables has been
developed since the pioneering work of \cite{Renyi62} for example developed in \cite{g78amm,Nagaraja98}.  
The applications in Physics started
somewhat later, but have by now found use in various problems related to
random walks, spin-glasses, type II superconductors and quantum
chaos. In
this work, our emphasis is on the record statistics in deterministic dynamical
systems, both classical and quantum.  In particular we choose the most studied
low-dimensional paradigm of Hamiltonian chaos, namely the standard map, or
known variously as the kicked-rotor, Chirikov-Taylor map etc.. Its
quantization has also been extensively studied, as well as realized
experimentally in cold atom set ups in \cite{MooRobBhaSunRai1995}.

Now to define records more precisely, given that $\{X_t, t=1, \cdots, N\}$
is a
finite time series, the first element, $R(1)$, of the corresponding records
series is $X_1$ itself and at subsequent times $t$ it will be
$R(t)=\mbox{max}(X_t,R(t-1))$. As $X_t$ is a random variable, so is $R(t)$ and
properties of this random variable is of interest. Thus $R(t)$ which consists
of elements of $X_t$ in a non-decreasing order are the upper record
sequence.
If the minimum value is taken instead of the maximum the record sequence is
called the lower record sequence.  From the definition of records, it is clear
that last upper (lower) record is also the global maximum (minimum) of the
sequence $\{X_j\}$. Hence, the statistics of last record will correspond to
similar results of maximum or minimum from extreme value theory. A word of
caution is warranted here, as the second last upper (lower) record will {\sl
not} be the second maximum (minimum) of the sequence. This can be easily
understood by visualizing a sequence $X_t$ whose global maximum (minimum)
occurs before second maximum (minimum). As in that case, record statistics
will not sense the presence of second maximum (minimum). To develop more
familiarity with record sequence, let's take $\sin x,~0\leq x\leq2 \pi$ as an
example. In the range $[0,\pi/2]$ the upper record sequence is $\sin x$ itself
and beyond that the entries will be constant 1.

Our motivations in this study are many. Firstly, while there is an elegant and
simple theory of records for the case of independent random variables, we can
naturally expect interesting departures for correlated processes. One such
class of problems, namely random walks, have been studied in this context and
remarkable departures and universalities have been uncovered. One of our
motivations is to see how much of this survives for deterministic dynamical
systems. For instance it has been appreciated now for more than 40 years that
deterministic systems such as the kicked rotor can display normal diffusion,
as well as anomalous diffusion. It is then of interest to ask if the record
statistics behaves in the same manner as that of random walks in this
scenario, and what happens in the different diffusion regimes. Also related
question is how do the record statistics change as the system undergoes a
transition to chaos. The latter question is natural here, but not so in the
setting of random walks. We also wish to go on to explore quantum dynamics and
review some of our recent works on record statistics. In this publication, all
the material pertaining to record statistics in the classical standard map or
kicked rotor is new, so also detailed derivations of the record
statistics for delta correlated variables, as well the lower record
statistics. Some other material, having to do with eigenfunction
statistics  have been partially presented in publications before (see
\cite{slj12arxiv}) and
is included here as a brief review.

\subsection{Independent and identically distributed ({\it i.i.d.}) random variables}
Consider the entries of $\{X_j\}$ as being independent and identically distributed
random numbers. As we have exchangeable entries, the symmetry of the
sequence
determines the probability of $X_j$ being a record to be $1/j$ at the $j^{th}$
trial; see \cite{Renyi62} or \cite{Nagaraja98}. In other words, any of the $j$ entries till
$X_j$, including itself, is equally possible to be the record. Consider
temperatures in a particular city, despite all the fluctuations, it is clear
that it beats its own record. In formal language, the process of setting a
record is persistent. Let us denote the number of records in a sequence of
length $n$ by $N_n$, then the above statement can be rephrased as $N_n
\rightarrow \infty$ with length of sequence $n \rightarrow \infty$.

The next natural question is about the frequency with which records are broken.
Define an indicator function
\begin{equation}\label{reco:eq:indicator}
I_j = \left\{ \begin{array}{ll}
 1& \textrm{if there is a record at $j$ },\\
0 & \textrm{otherwise}.
\end{array} \right.
\end{equation}
Calculating the expected value of $I_j$ is equivalent to calculating
the probability of $I_j$ taking the value 1, but this is precisely the
probability of $X_j$ being the record- which is $1/j$. Thus $\br I_j
\kt =1/j$.  Similarly, for variance, we note that expected value of
$I_j^2$ is the same as expected value of $I_j$. This immediately gives
the variance as $1/j - 1/j^2$.  It can be easily proven that $I_j$ are pairwise
uncorrelated (statistically independent).  
In other words the probability of the position of the records is a
Bernoulli process, $\mbox{Ber}(1/j)$.

As the total number of records $N_n$ in a sequence $\{X_1, X_2,
\cdots X_n\}$ is $\sum_{j=1}^n I_j$, the expectation and variance can
be readily calculated:
\beq\label{reco:eq:NRave}
\begin{aligned}
\br N_n \kt &= \sum_{j=1}^n \br I_j \kt = \sum_{j=1}^n\frac{1}{j} =
H_n. \\
   V(N_n)   &= \sum_{j=1}^n V(I_j) =
\sum_{j=1}^n\frac{1}{j}-\sum_{j=1}^n\frac{1}{j^2}.
\end{aligned}
\eeq
Here $H_n$ is the $n^{th}$ harmonic number.
A remarkable, well-known fact from the theory of records is that for
{\it i.i.d.} variables these quantities are {\it distribution-free},
that is
independent of the particular underlying distribution $p(x)$, see
\cite{Nagaraja98}. For example the average number of records $\br N_n\kt =
H_n\sim \log(n)+ \gamma$, where $\gamma$ is the Euler constant, is indeed very small compared to the length $n$ of
the data set; typically records are rare events in independent processes.
From the above it immediately follows that the variance also grows as
$\log(n)$.  Next section on, we will change the notation from $N_n$ to $N_R$.

\subsection{Random walks and records}

Some of the recent works concerns the behavior of records in
correlated processes, for example see \cite{mz08prl, wbk11pre, SchMaj2013}.  One
important class of such 
processes are random walks and the few
studies on their record statistics is now very briefly reviewed.
In case of random walks, $x_t$ represents the position of a random
walker at a time step, $t$.  For discrete time steps and jump-lengths
drawn from an {\it i.i.d.} symmetric distribution, Majumdar and Ziff
showed that the probability
$P(M,N)$ of $M$ records in $t=N$ steps is given by 
\beq
P(M,N)=\binom{2N-M+1}{N}2^{-2N+M-1},\quad M\leq N+1 
\eeq 
with mean $\br M\kt \sim \frac{2}{\sqrt{\pi}}\sqrt{N}$. Age statistics of record {\it i.e.},
how long a record survives before it is broken, is given by $\br l\kt \sim
\frac{N}{\br M\kt} \sim \sqrt{\frac{\pi N}{4}}$ (\cite{mz08prl}). Thus in strong
contrast to the {\it i.i.d.} case, the number of records is very large.

This makes physical sense, as the random walker is indeed trying to go somewhere and 
each time she has genuinely progressed in a particular direction a record is set. Being diffusive,
this process is certainly slower than ballistic, and in fact grows in the same way as the distance 
from the origin. That said, it is then interesting and somewhat surprising that it retains the property of being universal in that it is independent of the exact distribution of the jump-lengths.  In fact even when
distributions have a fat tail (like Cauchy distribution), this result
has been shown to be valid by \cite{mz08prl}.

 In the case of continuous time random walks, {\it i.e.} when position of a
random walker is now observed at equal discrete time steps with time interval
$\tau_0$, the probability $P(M,t)$ of finding $M$ records within a
given time duration
$t$, as derived by \cite{s11epl}, and attains a scaling form
given by,
\beq
\begin{aligned} P(M,t)&\sim (t/\tau_0)^{-\alpha/2}
g_\alpha\left(M(t/\tau_0)^{-\alpha/2}\right)\\ g_\alpha(x) &=
\frac{2}{\alpha}x^{-(1+2/\alpha)}L_{\alpha/2}(x^{-2/\alpha}), \quad
0<\alpha\leq1,
\end{aligned}
\eeq
where $L_{\alpha/2}(x)$ is a one-sided L\'evy stable
probability distribution function. The moments of records,
asymptotically go as
\beq
\br M^\nu\kt \sim \frac{(2/\alpha)\Gamma(\nu)}{\Gamma(\nu\alpha/2)}
(t/\tau_0)^{\nu\alpha/2}.
\eeq
The mean-age of the record is
\beq
\br l\kt=\br t/M\kt\sim \frac{\tau_0(\alpha/2)}{\Gamma(1-\alpha/2)}[\ln
(t/\tau_0)-\Psi(1-\alpha/2)](t/\tau_0)^{(1-\alpha)},
\eeq
where $\Psi(x)$ is di-gamma function, see \cite{s11epl}. The
parameter
$\alpha\in (0,1)$, appearing as exponent of power law decay of
the waiting time
distribution decides the qualitative behavior of record statistics.

 In another generalization due to \cite{wbk11pre},  the
record distribution for a random walk with discrete time steps but asymmetric
jump distribution, has been calculated for the model governed by
$x_n=x_{n-1}+\xi_n + c$ with $\xi_n$ being the jump length with symmetric
distribution while $c$ is the constant drift. For Gaussian jump distribution
with variance $\sigma$, in the limit of small drift (smallness is compared
with respect to $c=0$ case) {\it i.e.}  $\left(\frac{c}{\sigma} \ll
\frac{1}{\sqrt{n}}\right)$, the mean number of records and record rate ({\it
i.e.} probability of $n^{th}$ event being a record), is given by,
\beq
\begin{aligned}
\br M_n(c) \kt &\approx
\frac{2\sqrt{n}}{\sqrt{\pi}}+\frac{c}{\sigma}\frac{\sqrt{2}}{\pi} (n
\arctan(\sqrt{n})-\sqrt{n}),\\ P_n(c) &\approx \frac{1}{\sqrt{\pi
n}}\frac{c}{\sigma}\frac{\sqrt{2}}{\pi} \arctan(\sqrt{n}).
\end{aligned} \eeq
In the large drift limit, $P_n(c)$ approaches a constant
value \cite{wbk11pre}.  Later, this problem was exactly solved for
arbitrary value of constant bias $c$ and a number of interesting
results have been obtained such as the qualitatively different regions
in the space of $c,\mu$ by \cite{MajSchwer2012} where $\mu$ is the L\'evy index for symmetric
stable law of jump distribution.  

\subsection{Application of record statistics in superconductors and
  quantum chaos}
The geometric feature of random systems such as size of the largest
cluster in percolation on a finite lattice of size $N$, has been
shown to follow Gumbel distribution in the large-$N$ limit by
\cite{b00pre}. In type-II superconductors, temperature independence
of the
magnetic creep rate {\it i.e.} rate of change of magnetic field in the
sample at a constant temperature for a range of temperatures has been
understood in terms of record dynamics, namely, the dynamical
properties of the times at which a stochastic fluctuating signal (in
this case thermal noise due to non-zero temperature) establishes
records. This puzzling temperature independence of the creep rate,
which at non-zero temperatures has its origin due to thermal
fluctuations, has been sorted out by showing that the process of
vortex penetration into the sample can be described in terms of a
Poisson process with logarithmic time argument, called the log-Poisson
process; a result from record dynamics by \cite{ojns05prb}. 

The
application of record statistics in case of quantum chaos has been
treated in \cite{slj12arxiv}.
It is known that the eigenstate
intensities in fully chaotic systems with no particular symmetries
are conjectured to behave exactly as these random vectors subject only
to a normalization constraint. These are also the statistical
properties of eigenvectors of the classical ensembles of  random
matrix theory.  For chaotic systems, the applicability of  random
matrix theory \cite{Mehta04,brodypandeyrmp81} has been well
appreciated for long \cite{bgs84prl}.  As we will discuss in detail in
Section \ref{rec0:smintro} for the standard map the breaking of
the last KAM torus allows for diffusion in
phase space and diffusion being connected with random walks, we expect
and indeed observe the average number of records in eigenvectors to go
as $\sqrt{N}$. In fact this is first known instance where $K$
values at which the last KAM torus breaks has been seen in a quantity
derived from quantum mechanical spectrum.

\section{Records in $ \delta$-correlated variables}\label{sec:rec:delseq}
\subsection{Upper records for complex random vectors}
For a correlated sequence, let the probability density for a record
variable $R$, at time $t$ be $P(R,t)$. The average record is given by
$\left \langle R \right \rangle = \int \diff R ~R P(R,t)$.
Let $P(x_1,\ldots, x_N)$ be the j.p.d.f. of $N$ random variables. The
probability that the record at time $t$, is less than $R$ is
given by (for example \cite{slj12arxiv}):
\begin{equation}
Q(R,t)= \int_{0}^{R} \diff x_1 \cdots \diff x_t P_t(x_1, ...x_t)
\end{equation}
where $P_t(x_1,\ldots,x_t)= \int P(x_1, \ldots,x_N) \diff
x_{t+1}\cdots \diff x_{N}$ is the marginal j.p.d.f. of the first $t$
random variables .  It follows that
$P(R,t) =\diff Q(R,t)/\diff R$.

Let us specialize to the case of  $\delta$ correlated random variables {\it
i.e.} sum of the random entries is a constant. For example, components of
normalized complex random vectors $z_n=\br n |\psi\kt$, have the j.p.d.f:
\beq\label{eq:rec:jpdfZ}
P(z_1,z_2, \ldots, z_N) = (\Gamma(N)/\pi^N) \delta \left(\sum_{j=1}^N
|z_j|^2-1\right).
\eeq
This is also the distribution of the components of the eigenvectors
of the GUE or CUE (Gaussian or Circular unitary ensembles) random
matrices, for example see the book by \cite{Haake91}. It is the invariant uniform distribution under
an arbitrary
unitary transformations on the $2N-1$ dimensional sphere. It is the
unique (Haar) measure on $S^{2N-1}$. The normalization provides
correlation among the components that becomes weak for large $N$. The
intensities $x_i=|z_i|^2$ being the random variables of interest it
is more useful to define the j.p.d.f. directly in terms of these:
\begin{equation}\label{reco:eq:jpdfgue}
P(x_1,\dots, x_N;u) = \Gamma(N)\delta\pa{\sum_{i=1}^N x_i -u},
\end{equation}
where $u$ is an auxiliary quantity, the actual j.p.d.f. corresponding to
$u=1$.

The delta constraint also arises in problems such as the ``broken-stick" one,
 wherein a
stick is broken into a fixed number of pieces but with random lengths.  The
delta correlation is a very weak constraint, and in fact it is not hard to
conclude on some reflection that the number of records and its statistics
remains unaffected by this correlation. The way in which a set of random
numbers can be correlated by the delta function, gives us a hint: take {\it
i.i.d} random normal variables and normalize them, then the resulting set of
numbers are uniformly distributed on some sphere and hence are delta
correlated. But the rank order remains the same on a normalization process and
hence the record statistics remains the same. Note that we have here the new
variables $x_i$ being distributed uniformly on a simplex rather than a sphere:
however this does not matter for the records as again the act of taking a
square does not affect the rank order. Nevertheless we show this explicitly
here, as we derive the probability of the value of the record itself, which {\it does} differ from the uncorrelated case.
Also our analysis generalizes in a significant way
an earlier analysis of extreme values in such random variables by \cite{ltbm08prl}.

Defining
\beq
Q(R,t;u)=\int_{0}^{R} \diff x_1 \cdots \diff x_t \int_0^{\infty}
P(x_1,\dots, x_N;u) \diff x_{t+1}\cdots \diff x_{N}
\eeq
leads to
\beq
\int_0^{\infty} e^{-su} Q(R,t;u) \diff u =\frac{\Gamma(N)}{s^{N}}
\sum_{m=0}^t (-1)^m \binom{t}{m} e^{-s mR} .
\eeq
Using the convolution theorem, and then setting $u=1$ in $Q(R,t;u)$
gives
\begin{equation}\label{reco:eq:qrt}
Q(R,t) = \sum_{m=0}^t (-1)^m \binom{t}{m} (1 - mR)^{N-1} \Theta
(1-mR).
\end{equation}
Hence
\beq
P(R,t) = \sum_{m=1}^t (-1)^{m+1} \binom{t}{m} m (N-1)(1 - mR)^{N-2}
\Theta (1-mR),
\eeq
the probability density that the record is $R$ at time $t$. Note that
$P(R,N)$ is the
probability density that the maximum value of the entire data set is
$R$, which was calculated for the case of random GUE vectors in
\cite{ltbm08prl} and therefore $P(R,t)$ here is a  generalization.
The piecewise smooth probability distribution found there has a
similar behavior here.

 It was shown in \cite{ltbm08prl} that $P(R,N)$ is Gumbel distributed
asymptotically. In fact the generalization presented in Eq.
(\ref{reco:eq:qrt})  is also Gumbel distributed for large $N$, as for
large $N$ and large $t\gg 1$
\beq
Q(R,t) \approx (1-\exp(-NR))^t \approx \exp\left(-t \exp(-NR) \right).
\eeq
Since the Gumbel distribution is of the form
$\exp[-\exp(-(x-a_N)/b_N]$ where $a_N$ and $b_N$
are the shift and scaling. It follows that for the record statistics
the relevant parameters are
$a_N=\log(t)/N$ and $b_N=1/N$. The shift generalizes from $\log(N)/N$
for the maximum,
while the scaling remains the same.  The above form also appears in
the limit when the correlations are ignored.

The average value of the record as a function of time is
\beq \label{reco:eq:avRgue_exact}
\langle R(t)\rangle=1-\int_0^1 Q(R,t)\, dR=\frac{1}{N}\sum_{m=1}^t
(-1)^{m+1}  \frac{1}{m} \binom{t}{m} =
\frac{H_t}{N}=\frac{1}{N}\sum_{k=1}^t \frac{1}{k},
\eeq
where $H_t$ is a Harmonic number as defined above. Known asymptotics
of the Harmonic numbers implies that
\begin{equation}
\label{randomavgrecord}
\br R(t) \kt  = \frac{1}{N}\left( \gamma+ \ln(t) + \frac{1}{2t} -
\sum_{k=1}^{\infty} \frac{B_{2k}}{2k\,  t^{2k}} \right),
\end{equation}
where $B_{2k}$ are Bernoulli numbers, and $\gamma$ is the
Euler-Mascheroni constant. Again, this presents a generalization of
the average maximum intensity found in \cite{ltbm08prl} which
corresponds to $t=N$.

It is not hard to prove that for intensities of random states too,
the probability of the position of the records is a Bernoulli process
although they are correlated by the normalization constraint. Let
there be records at positions $(j_1=1 < j_2 < \cdots < j_m)$ and let
$I_{J_k}=1$ if there is a record at $j_k$ or $0$ otherwise. Then the
j.p.d.f.
\beq
\mbox{Prob}(I_{j_1}=1, I_{j_2}=1, \ldots, I_{j_m}=1)=\int_{{\cal C}}
P(x_1, \ldots, x_N; u=1) dx_1\cdots dx_N= \prod_{k=1}^m\frac{1}{j_k}.
\eeq
Here ${\cal C}$ is the set of constraints: $0\le x_k \le x_{j_2},$
$j_1 \le k \le j_2-1$; $0 \le x_k \le x_{j_3},$ $j_2 \le k \le
j_3-1$; $\cdots$, $0 \le x_k \le x_{j_m},$ $j_{m-1} \le k \le j_m-1$;
$0 \le x_k \le 1$, $j_m \le  k \le N$. The above result follows on
using the Laplace transform to free the constraint in Eq.
 (\ref{reco:eq:jpdfgue}). However this is the result for {\it i.i.d.} random
variables, and implies that the occurrence of a record at $j_k$ is an
independent process, as the above is valid for all arbitrary choices
of the locations $j_k$. Hence $\mbox{Prob}(I_j=1)=1/j$ and $
\mbox{Prob}(I_j=0)=1-1/j$, in other words the process is
$\mbox{Ber}(1/j)$.

The average number of records is thus
\beq
\br N_R \kt = \left \br \sum_{j=1}^N I_j \right \kt  = \sum_{j=1}^N
\frac{1}{j}=H_N,
\eeq
while as a random variable $N_R$  has a distribution essentially
given by the signless Stirling numbers of the first kind, some times
called the Karamata-Stirling law (see \cite{Nev87tpa}). Such laws hold for
a variety of disparate processes including the number of cycles in a
random permutation of $N$ objects, number of nodes in extreme side
branch of random binary search trees etc. \cite{bhl98rsa}. Being
distribution-free, the number of records is a statistics that
directly detects correlations.

The probability that the final record, which is the maximum in the
entire data sequence, lasts
for time $m$ can also be simply calculated:
denoted $S_N(m)=P(I_N=0,I_{N-1}=0, \cdots, I_{N-m+2}=0,
I_{N-m+1}=1)=1/N$, it is (somewhat surprisingly) independent of $m$,
and uniform.  This implies that the {\it position} at which the
maximum occurs is uniformly distributed. The implications of this for
quantum chaotic wavefunctions where strong scarring effects of
classical periodic orbits ( see \cite{mk79prl,HellerPRL84, bj90pra,
sridhar95prl,laurent07prl}) can affect the maxima of states is of
natural interest, and will be discussed below.

\subsection{Lower records in complex random vectors}
The question of a record minimum can be asked in a similar way, the
cumulative density function for a record minimum, {\it i.e.} the
probability that the record is ``greater'' than $R$
at a time $t$ is given by
\begin{equation}
Q(R,t)= \int_{R}^{\infty} \diff x_1 \cdots \diff x_t P_t(x_1,
\dots,x_t)
\end{equation}
where $P(x_1,\dots, x_N)$ is j.p.d.f. of $N$ random variables, and
$P_t(x_1,\dots, x_t)$ is the j.p.d.f. of $t$ random variables given by
$\int_{all~ range} P(x_1,\dots,x_N)   \diff x_{t+1} \dots \diff
x_{N}$.
From $Q(R,t)$ we can get $P(R,t)$ as
\begin{equation}
P(R,t) =-\frac{\diff Q(R,t)}{\diff R}.
\end{equation}

It is not difficult to calculate the probability distribution of
record minima in the case of a complex random vector, essentially
following the same technique as in the case of maximum records,
\begin{eqnarray}
\nonumber
Q(R,t;s)&=&\Gamma(N) e^{-s\sum_{i=1}^t
r_i}\left[\prod_{i=1}^t\int_{R}^\infty \diff r_i e^{-s r_i} \right]
\left[\prod_{i=t+1}^N\int_{0}^\infty \diff r_i e^{-s r_i} \right]\\
\nonumber
~ &=&\frac{\Gamma(N)}{s^{N-t}} \frac{e^{-sRt}}{s^t}
\end{eqnarray}
Using the convolution theorem and then replacing $u=1$, we
get
\begin{equation}\label{reco:eq:qrt_min}
Q(R,t) = (1 - R t)^{N-1} \Theta (1-R t)
\end{equation}
Density,
\begin{equation}
P(R,t) =-\frac{dQ(R,t)}{dR}=(N-1)t(1-Rt)^{N-2}\Theta (1-R t)
\end{equation}
and  average record as a function of $t$ can be easily found
\begin{eqnarray}\label{eq:rec:rt_tmin}
\nonumber
\langle R(t)\rangle &=& \int R P(R,t) \diff R \\ \nonumber
		    &=&\frac{1}{Nt}.
\end{eqnarray}
Again this generalizes the results obtained in \cite{ltbm08prl}. For
large $N$ limit $Q(R,t)$ is again exponential and for $t=N$, it
retrieves all the results obtained in \cite{ltbm08prl}. In particular while the average intensity is $1/N$,the average {\it minimum} intensity is $1/N^2$ corresponding to $t=N$. The record low intensity is inversely proportion to the ``time", which in the case of static wavefunctions is the index of the basis used. It is easy to
show (by just changing the conditions and hence the limits in
appropriate integrations) that these are also Bernoulli process,
$\mbox{Ber}\pa{1/n}$ and hence all the results for survival probability,
lifetime distribution {\it etc.} remains same in this case too. 
We remark that the exponential form of $Q(R,t)$ in the large $N$ limit is
a special case of the Weibull distribution of the standard theory of
extreme value statistics of {\it i.i.d.} random variables.

\section{Record statistics in the classical and quantum standard map} \label{rec0:smintro}
Attention is now turned to the standard map. This is chosen as
the standard map is a simply defined dynamical system which has a
well-studied transition to chaos through the usual route of smooth two-degree of freedom 
Hamiltonian systems.
 It also has a simple and well-studied quantization and
allows breaking parity and time-reversal symmetries through quantum phases
and hence allows for studying GUE, GOE, (or CUE, COE), as well as
intermediate statistics. Before discussing the record statistics for
eigenvectors we summarize some aspects of the standard map
pertaining to quantization and the intensity distribution of its
eigenvectors.

\subsection{Elementary aspects of the standard map}
We recall for convenience some well-known basic aspects of the standard map.
The Hamiltonian of the $\delta$-kicked rotor  is given by
\beq
H(q,p,t) = \frac{p^2}{2} - \frac{g}{4\pi^2}\cos(2\pi q)
\sum_{n=-\infty}^\infty \delta\pa{t/T-n}.
\eeq
On rescaling $Tp \rightarrow p$ and $T^2 g \rightarrow K$, this results in
the area-preserving map:
\beqa \label{reco:eq:stdmapdef}
\nonumber
q_{n+1} &=& (q_n + p_n)~ \mbox{mod} ~1, \\
p_{n+1} &=& p_n - \frac{K}{2\pi} \sin(2 \pi q_{n+1}),
\eeqa
connecting the position and momentum just after kick $n$ ($q_n,p_n)$
to that after kick $n+1$.
This is the {\it Chirikov-Taylor} map or {\it
standard map}. The fixed points $(0,0)$, $(1/2,0)$ of this map
are stable and unstable respectively for small $K$.

The periodic property of $q$ endows the phase space with the topology of a
cylinder. It is also well known that the translational boost in momentum by $n\in \mathbb{Z}$ is
a symmetry of the standard map, and this allows one to take the modulo 1
for momentum as well and hence endows the topology of the two-torus to the
phase space. Similarly, it also enjoys a discrete symmetry of
reflection about the centre of square {\it i.e. $p\rightarrow (1-p)$,
$q\rightarrow (1-q)$}.  In the limit $K=0$, this map is completely
integrable and with the increase in $K$, destruction and creation of
invariant surface ({\it i.e. KAM tori})
takes place.

About $K\approx 1$, the last rotationally invariant KAM torus breaks, allowing global
diffusion in the momentum space. If the standard map is unfolded to a
cylinder it displays normal diffusion in momentum for large enough
$K$. When $K \gg 5$, the classical map is essentially fully chaotic.
While for very small $K$  the phase space seems to constitute only invariant curves, the $K=10$ 
case appears to be completely hyperbolic with no traces of islands.
At $K \approx 1$ a well-known transition takes place when the last rotational KAM torus breaks and 
allows for global momentum transport when viewed on a cylindrical phase space.
When $K<5$ the phase space is clearly mixed with large stable islands and chaotic regions coexisting.

To study the signatures of chaos in quantum mechanical spectrum of
such systems has been the concern of the subject of ``quantum chaos''
or ``quantum chaology''. Naturally quantization of such maps were the first step
\cite{CasFor1979,BerBalTabVor1979}.  Analogous to Eq.
(\ref{reco:eq:stdmapdef}),
Heisenberg equations can be integrated to yield similar equation for the
operators. The unitary operator connecting states separated by a period of the
kick is the quantum map. For the standard map on the plane, it is easy to see
that corresponding unitary operator will be (\cite{ChaShi1986,Izr1990}),
\beq \label{reco:eq:chrec:unitary_plane}
\hat{U} = \exp\pa{-\frac{igT}{2\pi^2 \hbar}\cos(2\pi
\hat{q})}\exp\pa{-\frac{iT}{2\hbar}\hat{p}^2}.
\eeq
Due to periodicity in momentum and cyclic nature of the position
variable, the natural phase space will be $[0,1)\times [0,1)$ so this
map needs to be quantized on torus. The periodicity of both the
canonical variables implies that the Hilbert space of quantum
mechanics is finite dimensional. This dimensionality $N$ is related
to a scaled Planck constant as $N=1/h$ and hence the classical limit
is the large $N$ limit.
 A description can be found in the notes of  \cite{ArulSerc, Bae2003}.
 
The quantized standard map in position basis on
torus takes the form,
\beqa\label{reco:eq:chrec:uniop}
\nonumber
U_{nn'} &=&\frac{1}{N} \sum_{m=0}^{N-1} \exp\left[ -i \pi
\frac{(m+\beta)^2}{N} + 2 \pi i \frac{(m+\beta)}{N}(n-n') \right ]\\
 &~&\quad \quad \quad \quad \times \exp \left[ i \frac{K N }{2
\pi}\cos \frac{2\pi (n'+\alpha)}{N} \right].
\eeqa
$\beta$ and $\alpha$ are the phases which a state acquires along
position and momentum directions respectively. For periodic boundary
conditions,  $\beta=0$  while for anti-periodic, $\beta=1/2$.
For all other $\beta$ values, time-reversal symmetry is broken. A
similar role is played by $\alpha$ for parity symmetry.
We will focus  for the case, $\beta \ne 0$ and $\alpha \ne 0,1/2$
when we can expect that both the time-reversal symmetry and parity
symmetry are broken and the typical eigenstates would be like complex
random states.

\begin{figure}[htbp]
\centering
   \begin{subfigure}[t]{0.48\linewidth} \centering
     \begin{overpic}[width=2in]{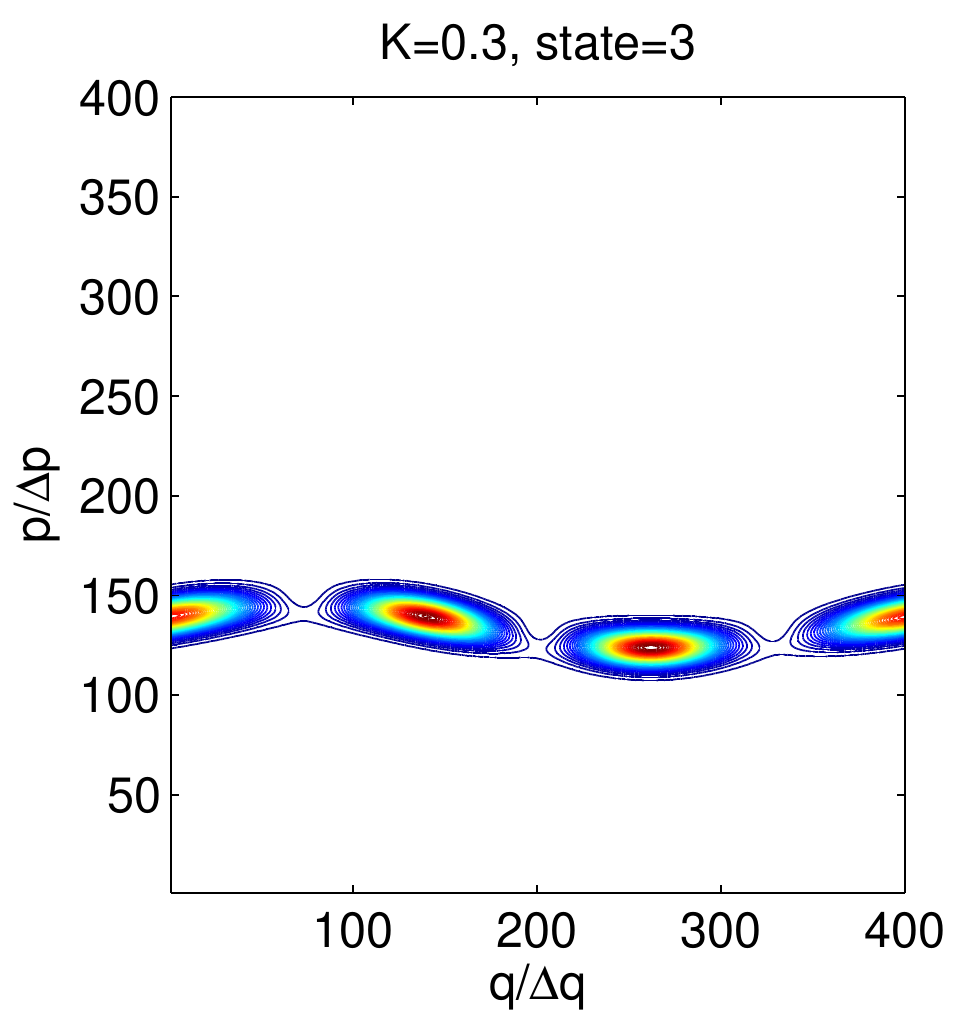}

\put(18,60){\includegraphics[width=1.5in]{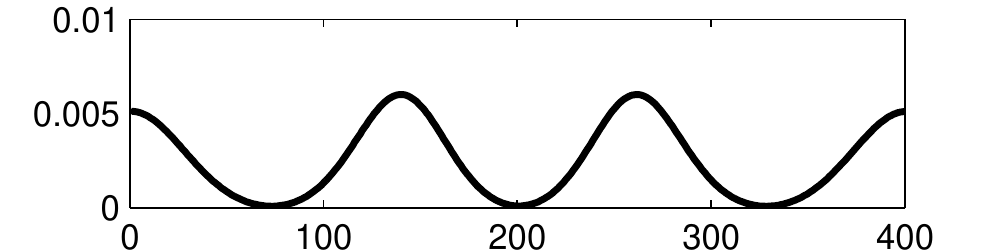}}
      \end{overpic}
     \caption{}\label{fig:chrec:inteK0p3a}
   \end{subfigure}
   \begin{subfigure}[t]{0.02\linewidth}\centering
   \end{subfigure}
   \begin{subfigure}[t]{0.48\linewidth} \centering
     \begin{overpic}[width=2in]{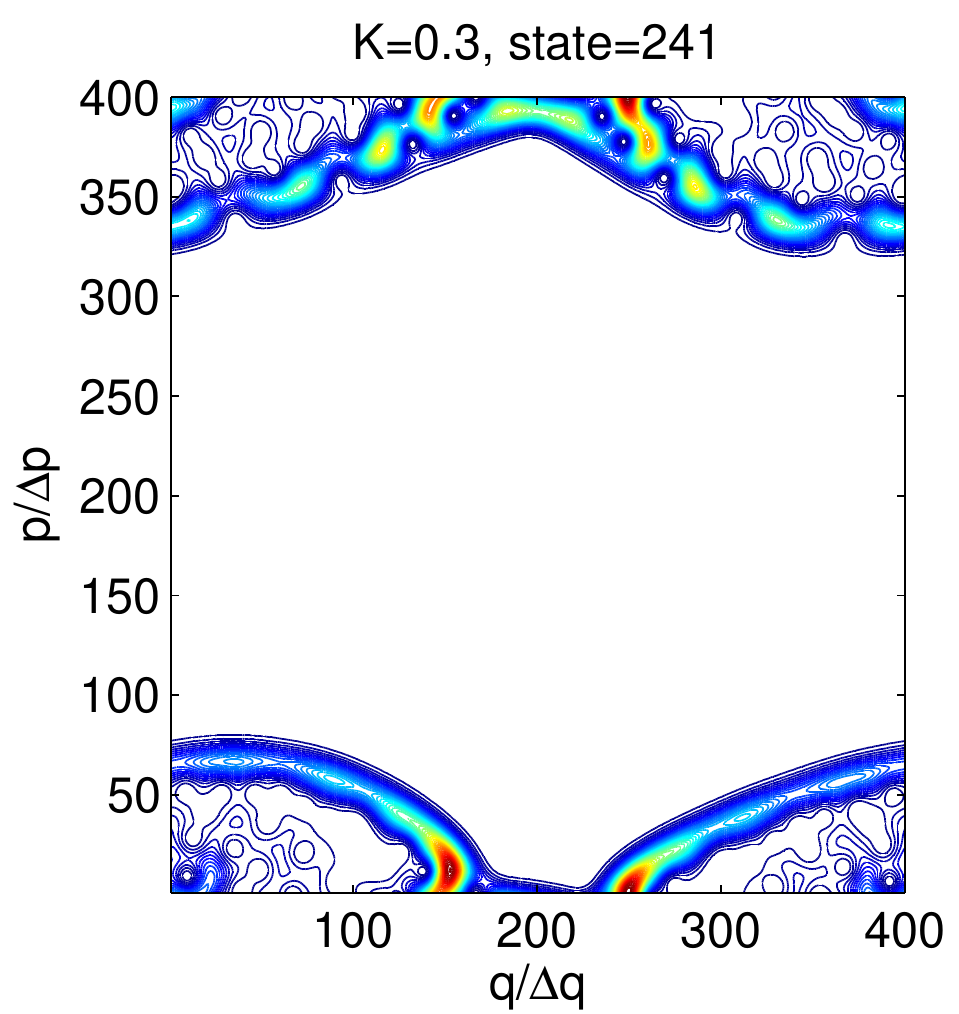}
        \put(18,50){\includegraphics[width=1.5in]{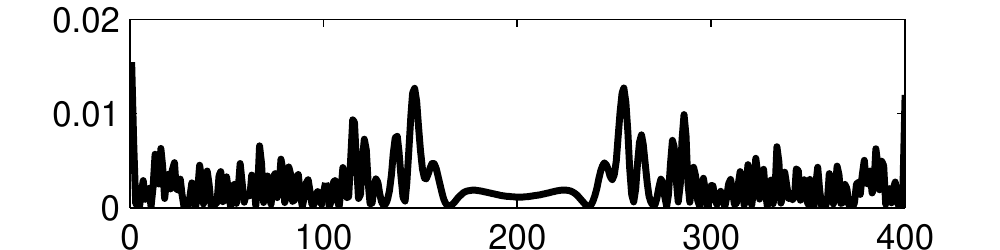}}
      \end{overpic}
     \caption{}\label{fig:chrec:inteK0p3b}
   \end{subfigure}
   \begin{subfigure}[t]{0.48\linewidth} \centering
     \begin{overpic}[width=2in]{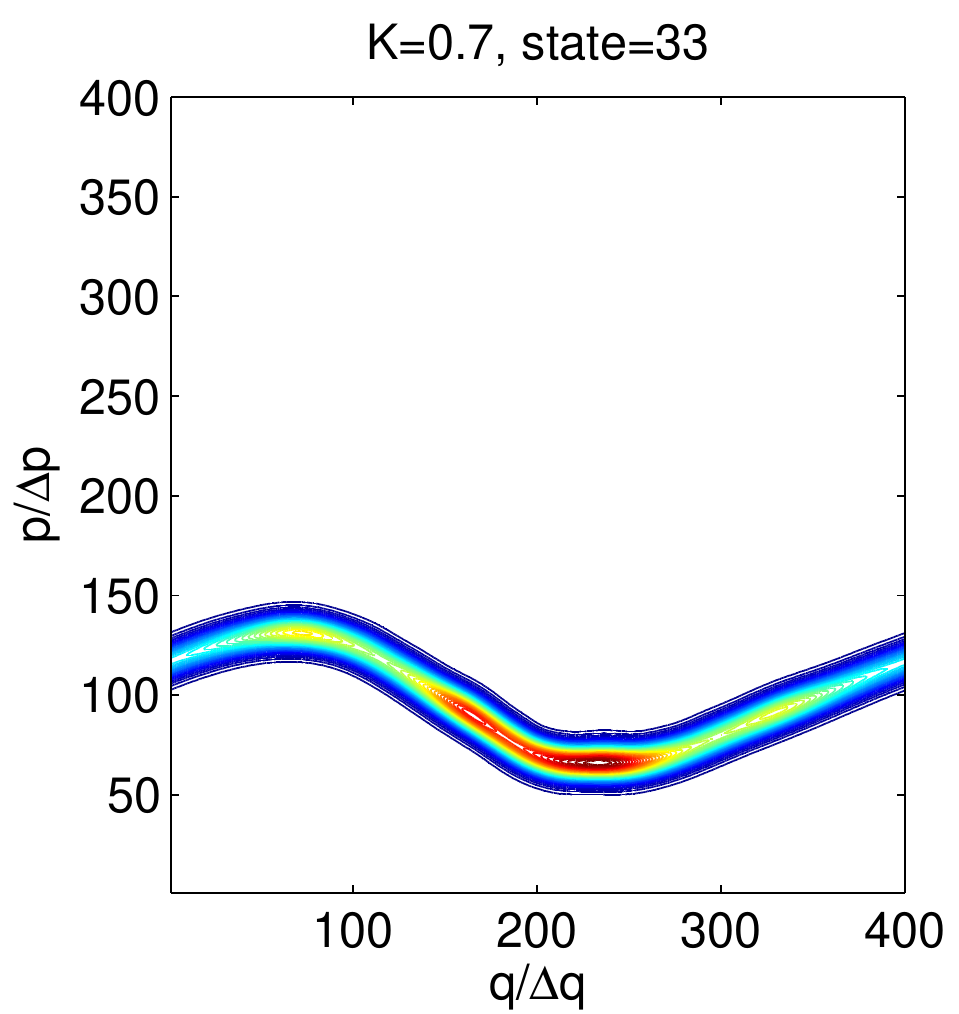}
        \put(18,60){\includegraphics[width=1.5in]{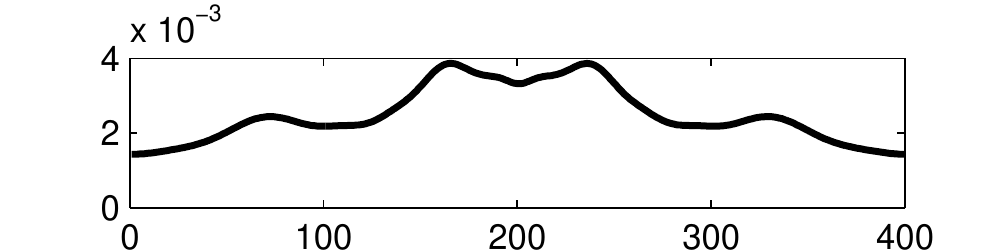}}
      \end{overpic}
     \caption{}\label{fig:chrec:inteK0p7a}
   \end{subfigure}
   \begin{subfigure}[t]{0.02\linewidth}\centering
   \end{subfigure}
   \begin{subfigure}[t]{0.48\linewidth} \centering
     \begin{overpic}[width=2in]{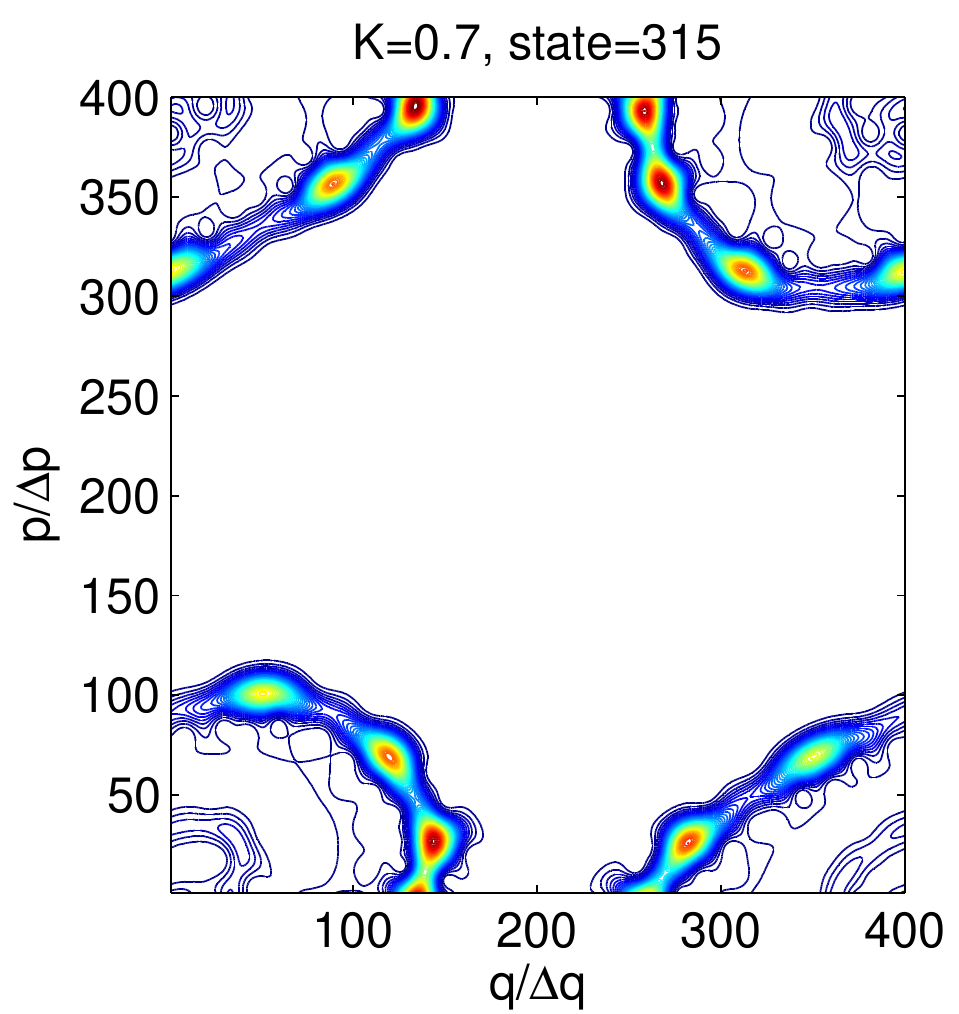}
        \put(18,50){\includegraphics[width=1.5in]{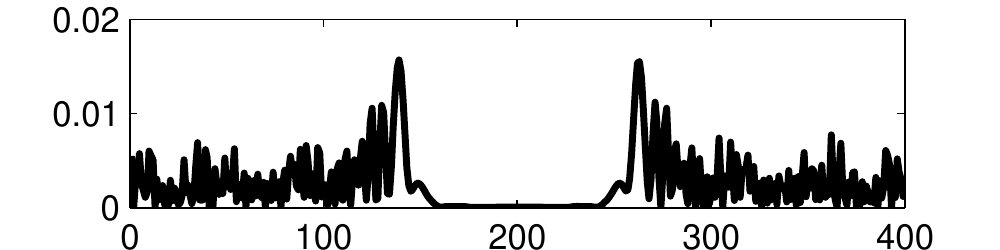}}
      \end{overpic}
     \caption{}\label{fig:chrec:inteK0p7b}
   \end{subfigure}
   \begin{subfigure}[t]{0.48\linewidth} \centering
     \begin{overpic}[width=2in]{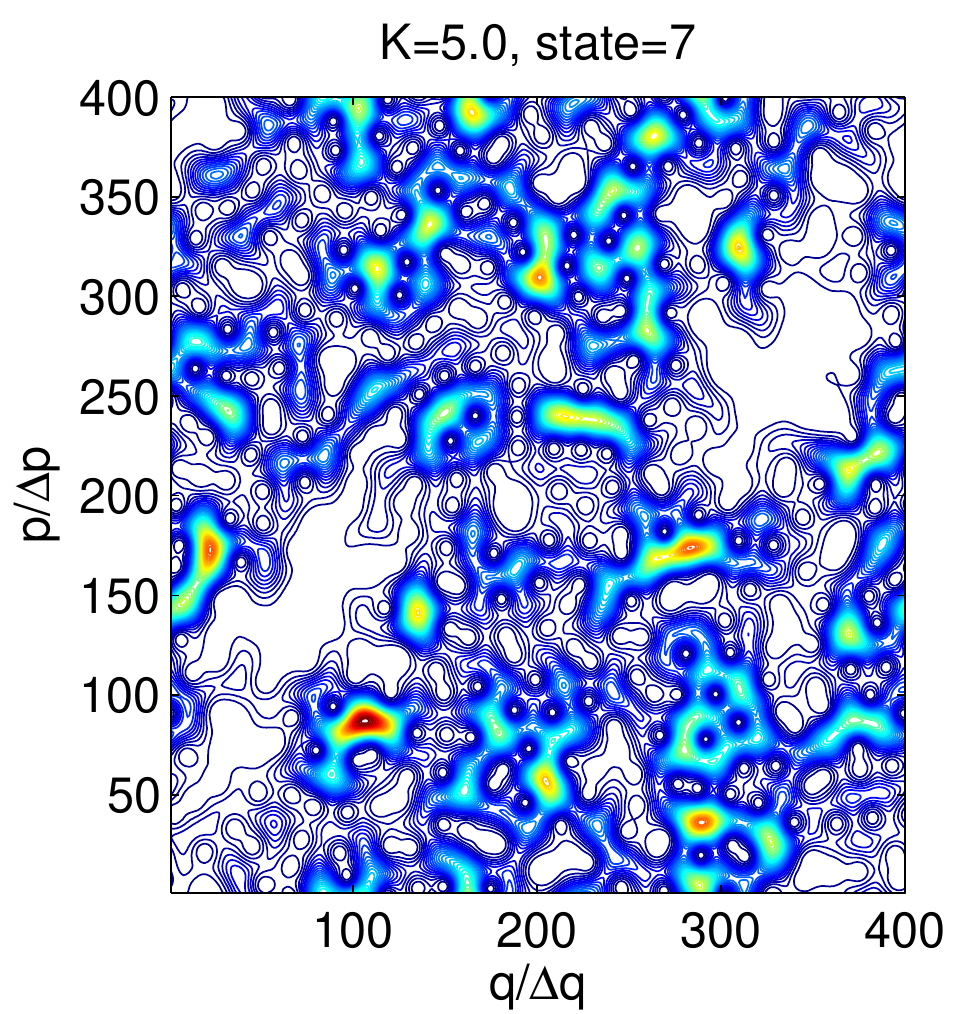}
        \put(18,60){\includegraphics[width=1.5in]{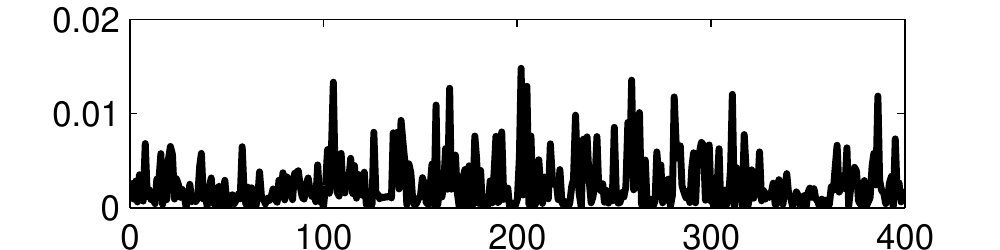}}
      \end{overpic}
     \caption{}\label{fig:chrec:inteK5a}
   \end{subfigure}
   \begin{subfigure}[t]{0.02\linewidth}\centering
   \end{subfigure}
   \begin{subfigure}[t]{0.48\linewidth} \centering
     \begin{overpic}[width=2in]{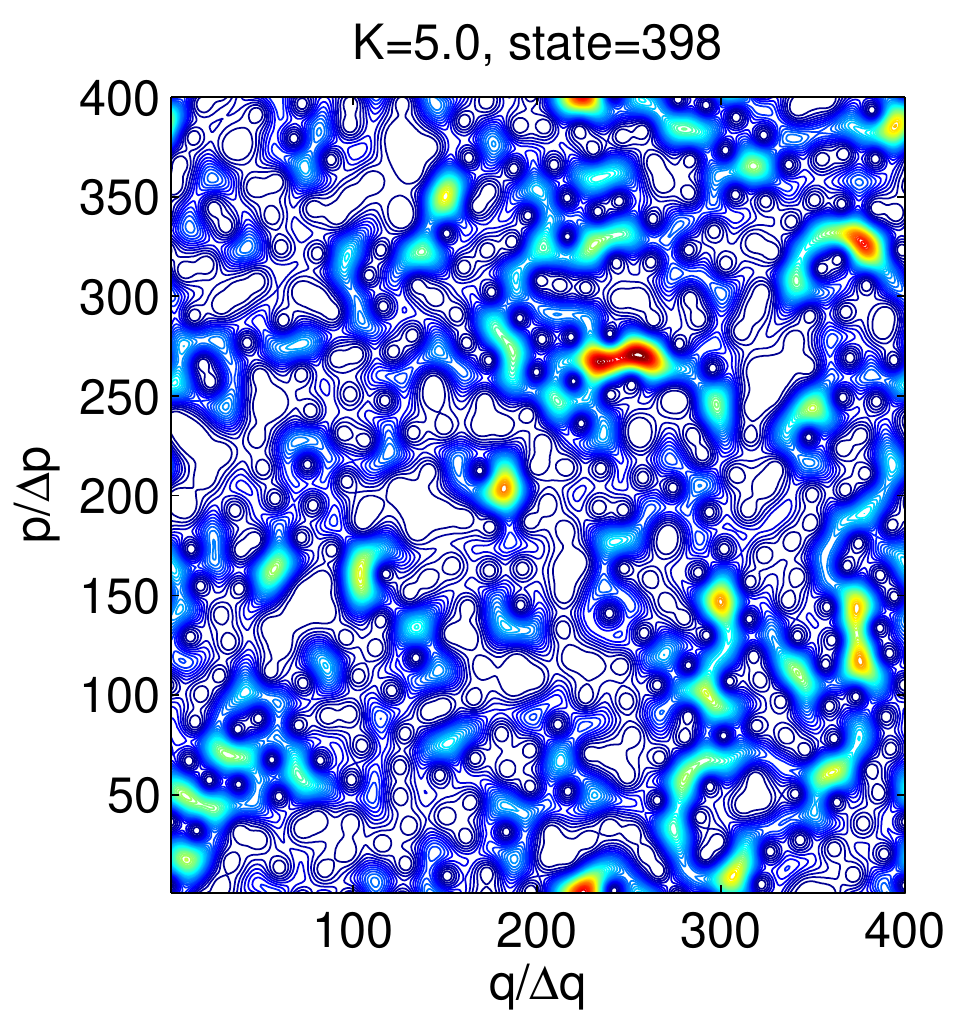}
        \put(18,60){\includegraphics[width=1.5in]{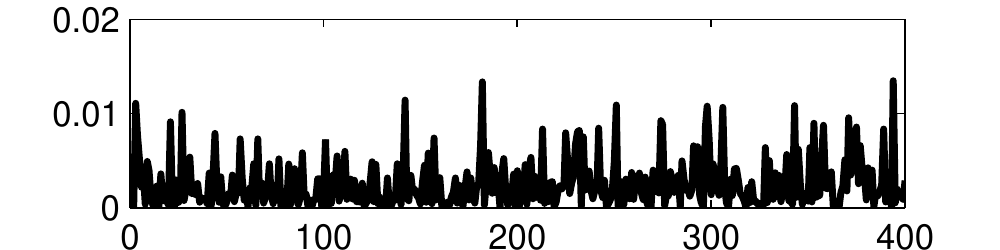}}
      \end{overpic}
     \caption{}\label{fig:chrec:inteK5b}
   \end{subfigure}
\caption{For different $K$ value the Husimi function, $|\br
q,p|\psi\kt|^2$,  of representative states and respective intensity
plots are shown where $|q, p\kt$ is coherent state which has been
calculated following reference \cite{saraceno_husimi}. Parity seems
to be hold good for lower $K$ values despite the value of $\alpha$
being 0.25 ({\it i.e. maximally broken parity}).}
\label{fig:chrec:inteK}
\end{figure}

To develop more familiarity with the intensities in different
eigenfunctions of the quantum standard map, we have plotted them in Fig.
\ref{fig:chrec:inteK} for $K = 0.3, 0.7$ and 5. It is clear that up
to $K=1$, intensities are (almost) symmetric about $p=1/2$ despite
$\alpha$ being $0.25$. This will have an impact when different
aspects of record statistics of intensity vectors are considered.

\subsection{Records and anomalous transport in the classical standard map}

The transport of momentum in the standard map has been studied extensively. It is known that in the strongly chaotic regimes the diffusion might be normal or
anomalous depending on the absence or presence of very special orbits, namely the accelerator modes.
Thus momentum diffuses either as normal
Brownian motion or if anomalous then as L\'evy walks
(\cite{Chi1979, MacMeiPer1984}).
Recently a detailed analysis on the
role of accelerator modes for anomalous diffusion has been carried
out using the so-called GALI (Generalized ALignment Index) method
 (\cite{ManRob2014}).  In this section,  numerical results for various
 quantities in the record
statistics for momentum transport in the standard map is presented. The time series
being studied are the squared deviation of momentum at $n^\text{th}$
iteration from the initial momentum.
As have been already mentioned in the introduction that it has been shown in \cite{mz08prl}, the average record for
random walk  or Levy walk (both), goes as $\sim \sqrt{4 t/\pi}$ with
time, $t$ while the variance of record depends on time linearly
{\it i.e.} as $2(1-2/\pi)t$. It is then of interest to ask if this is the case for the standard map as well.

For large $K$, when the standard map is strongly chaotic,
the distribution of position is uniform in range $[0,1)$. This
suggests that in chaotic regime one can take the position variable as
uncorrelated. Then, the standard map reduces to an effective one
dimensional stochastic map:
\beq
p_{n+1}=p_n-\f{K}{2\pi}\sin(2\pi r_n)~~\text{with} ~~r_n\in
\text{Uniform}[0,1)
\eeq
This is precisely the random walk with jump distribution taken from a
{\it i.i.d.} distribution. This clearly brings out not only why the
results for random walk can be expected to describe the results for record
statistics for standard map in the  hard chaos regime but also that
any departure from it is due to correlations present in the steps of the deterministic dynamical system.
For mixed phase regime, when a portion of
phase space remains quasi-integrable, the time evolution of average records
is expected to be sub-diffusive. But, any deviation in the large $K$ region can be a signature of accelerator modes.

\begin{figure}[h]
\begin{center}
\includegraphics[width=0.45\textwidth]{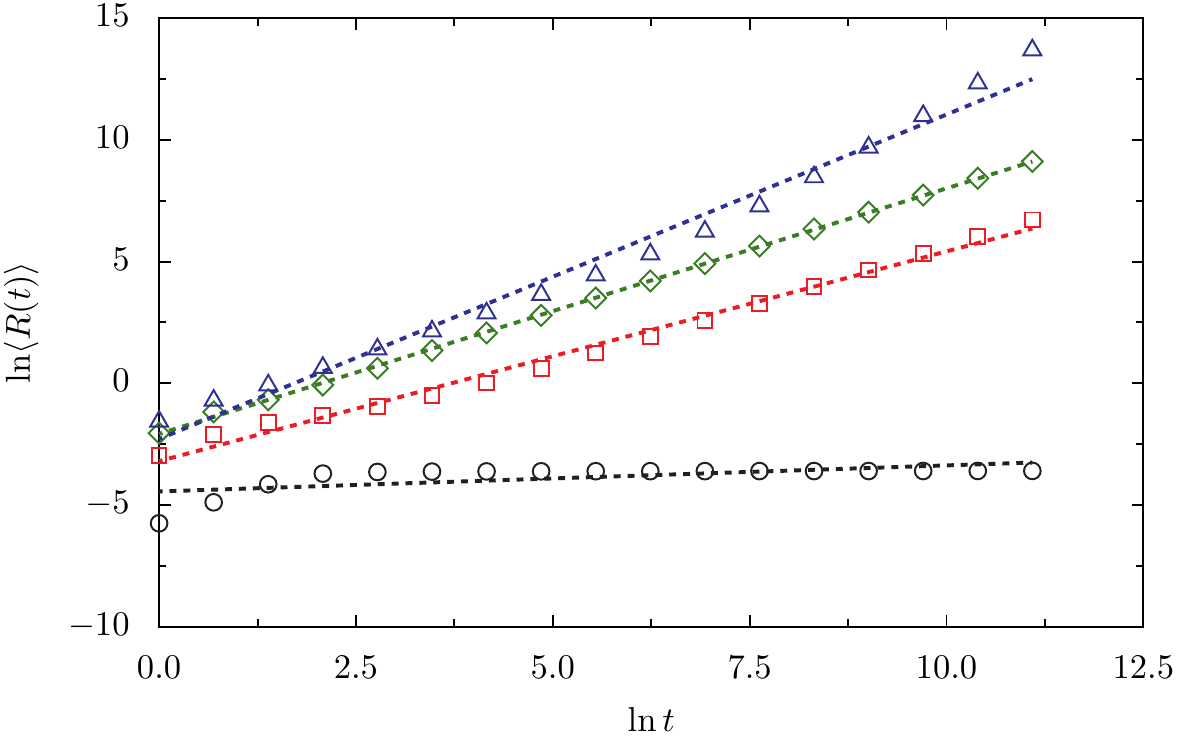}
\includegraphics[width=0.45\textwidth]{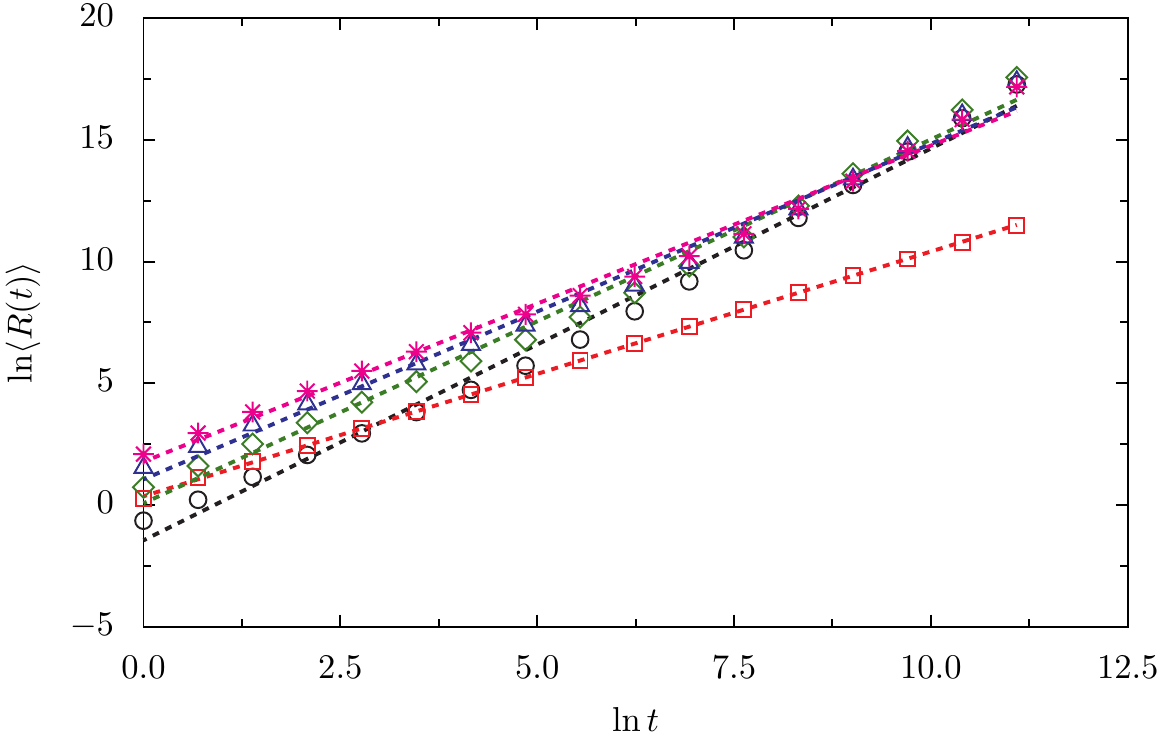}
\end{center}
\caption{ Average record of deviation in momentum square
as a
function of time has been plotted on log-log scale. Clearly the
long-time behaviour is $\propto t^\alpha$. {\bf (Left panel)} $K$
values and the fitted exponent in the power law, are $K=0.5$, with
$\alpha=0.11$ denoted by $\circ$, $K=2.00$, with $\alpha=0.86$
denoted by ${\color{red}\Box}$, $K=3.2$ with $\alpha=1.009$ denoted
by ${\color{green}\Diamond}$ and $K=4.05, {\color{blue} \triangle}$
with $\alpha=1.33$. It is known that $K=4.05$, there exist an
accelerator mode of period 2. {\bf (Right panel)} The exponent
$\alpha$ goes to $\sim 1$ for $K=10$ (denoted by ${\color{red}\Box}$)
i.e. in fully chaotic region. A slightly higher value of
$\alpha$ for all the $K$ corresponding to the regime where
accelerator modes are present in the phase space ($K=6.4$ with
$\alpha=1.61$ denoted by $\circ$, $K=12.70, {\color{green}\Diamond}$
with $\alpha=1.495$, $K=19.0, {\color{blue} \triangle}$ with
$\alpha=1.3788$ and $K=25.25, {\color{magenta}\ast}$ with
$\alpha=1.2992$). The dashed lines are the fitting curves. }
 \label{fig:rt_K}
\end{figure}

Rewriting the Eq. \ref{reco:eq:stdmapdef} as 
\beq
(p_n-p_0)^2 = \frac{K^2}{4\pi^2}\left(\sum_{i=0}^{n-1}  \sin^2(2\pi
q_i) + \sum_{i\neq j} \sin(2\pi q_i) \sin(2\pi q_j)\right).
\eeq
The record and the number of records for the quantity $(p_n-p_0)^2$,
namely the deviation in momentum square,  are
studied in this paper. Since, the square root of deviation in momentum
square, as argued previously, will behave like random walk, the exponent in
average record as a function of time will become 1 for true random
walk cases.  We have plotted the average record as a function of time in
Fig. \ref{fig:rt_K} and clearly the long-time behavior is a powerlaw $\propto
t^\alpha$. The averaging is done over 10000
trajectories with initial conditions chosen uniformly from
$[0,1)\times [0,1)$. The left panel corresponds to the case of mixed phase space or small $K$. We see the exponent increasing from values smaller than $1$ to $1$ and for $K=4.05$ be as high as $1.33$. At this value of $K$ there is still a mixed phase space regime, with the stable periodic orbit at $(0,0)$ having just undergone a bifurcation and having become unstable. This large value of the exponent may then arise from the presence of period-2 accelerator modes that have not nearly been studied as much as the usual period-1 modes.

\begin{figure}[h]
\begin{center}
\includegraphics[width=0.45\textwidth]
{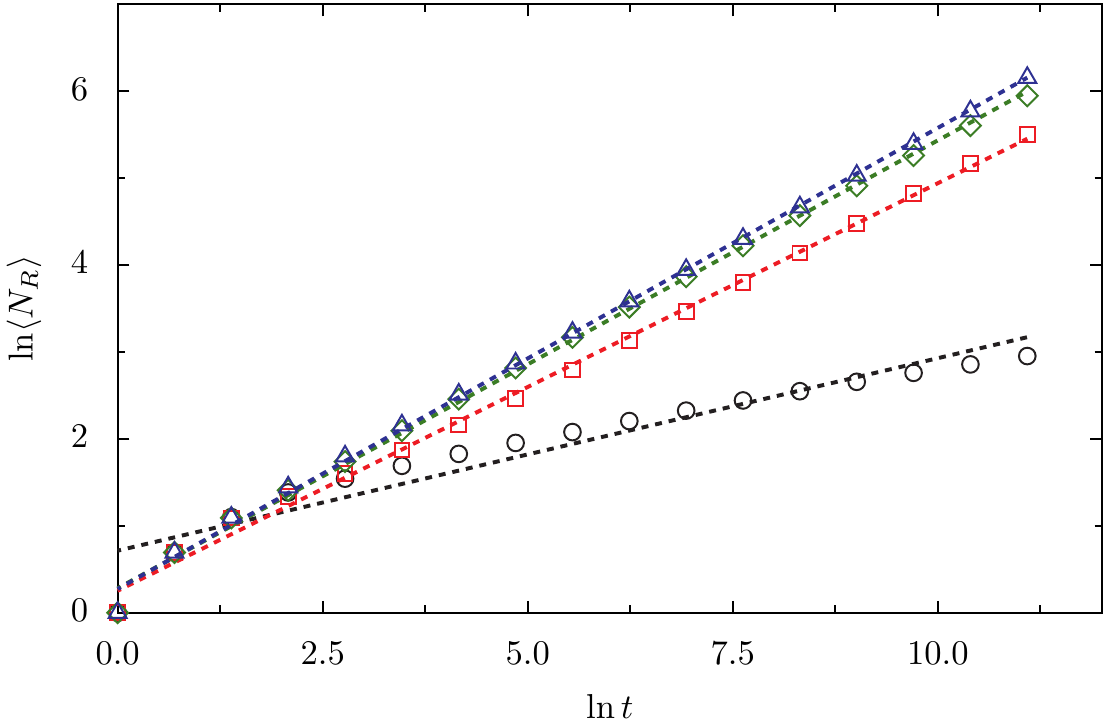}
\includegraphics[width=0.45\textwidth]{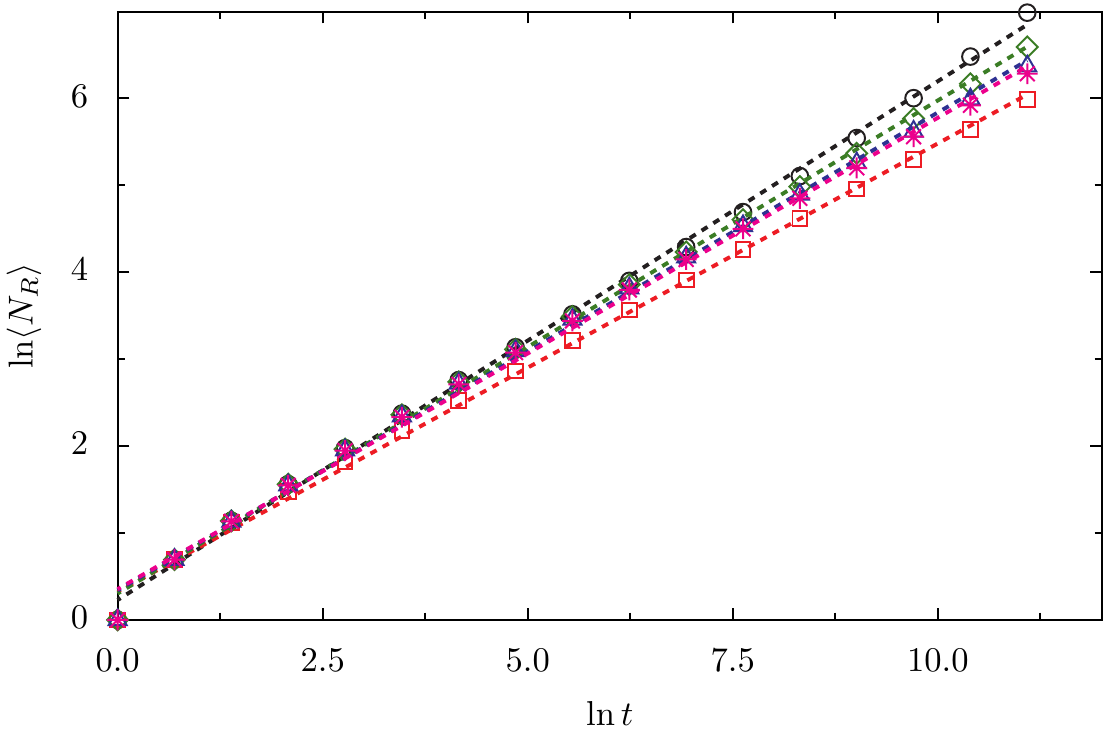}
\end{center}
\caption{ Average number of records as function of time has been
shown for various $K$ values. The averaging is done over 10000
trajectories with initial conditions for position drawn uniformly from
[0,1) and momentum from uniform distribution in $[0,1)$. {\bf (Left
panel)} $K$
values and the fitted exponent in the power law, are $K=0.5$, with
$\alpha=0.22$ denoted by $\circ$, $K=2.00$, with $\alpha=0.47$
denoted by ${\color{red}\Box}$, $K=3.2$ with $\alpha=0.51$ denoted
by ${\color{green}\Diamond}$ and $K=4.05, {\color{blue} \triangle}$
with $\alpha=0.53$.  {\bf (Right panel)} For hard
chaos regime ($K=10$ denoted by ${\color{red}\Box}$ with exponent
0.52) again the $\langle R_n\rangle$ varies in agreement with the
analytical results for random walk. In figure, the various K
values with the fitted straight lines  are, $K=6.4$ with exponent
0.60, denoted by $\circ$, $K=12.70, {\color{green}\Diamond}$
with exponent 0.57, $K=19.0, {\color{blue} \triangle}$ with exponent
0.55 and $K=25.25, {\color{magenta}\ast}$ with exponent 0.54. The
dashed lines are the fitting curves.The values quoted here are the
best fitted exponents.}
 \label{fig:rN_K_class}
\end{figure}

The right panel shows the case for $K \gg5$ or hard chaos regime. At
$K=10$ there is no evidence of any stable region or accelerator modes
(\cite{TomLak2007}) and it is found that $\alpha\sim
1.0$. However the same figure also shows several other large $K$
values that deviates from this and has larger exponents. In all these
values there are known accelerator modes. As we have averaged over
10000 trajectories
with initial momentum drawn from uniform distribution in $[0,1)$, a
few will satisfy the initial condition for accelerator modes which
explains why the $\langle R(t)\rangle$ for $(p_n-p_0)^2$ is not
increasing quadratically with
time.  This shows that simple record statistics can be used as a
signature to find the anomalous
transport region. This clearly captures the presence of accelerator
modes even of higher periods as can be seen in left panel of Fig.
\ref{fig:rt_K}.

In Fig. \ref{fig:rN_K_class}, we have plotted the
average {\it number} of records upto time $t$. Again the averaging is done
over the ensemble of 10000 trajectories with initial conditions chosen
uniformly from $[0,1)\times [0,1)$. The left panel again corresponds
to a mixed phase space regime, while the right to hard chaos. Once
again the same trends are observed and accelerator modes seem to have
a strong influence on the number of records set. While in the absence
of these special structures and for the case of hard chaos $N_R(t)
\sim \sqrt{t}$,
which is in agreement with the analytical results obtained in
\cite{mz08prl} for random walks. This implies that the average time for which records lasts 
in the classical standard map's diffusive regime also increases as $\sqrt{t}$ as in the random walk.
It will be interesting to study the maximum and minimum intervals over which records last, but we have 
not pursued that here. In the case of random walks this has recently been studied by  \cite{Godreche2014}.
 
However it should be pointed out that
even if the steps lengths were that of a Levy walk the scaling was
$\sqrt{t}$ for random walks, while in the case of a deterministic
dynamical system, the L\'evy walk leads to easily discernible
departures. Thus these L\'evy walks are evidently correlated ones and
the correlation plays an important role here. Analytical studies of
such walks is warranted in this context.

\subsection{Record statistics for standard map
eigenvector intensities}\label{reco:sec:rsStandmap}

\begin{figure}[htb]
\begin{center}
\includegraphics[scale=0.4]{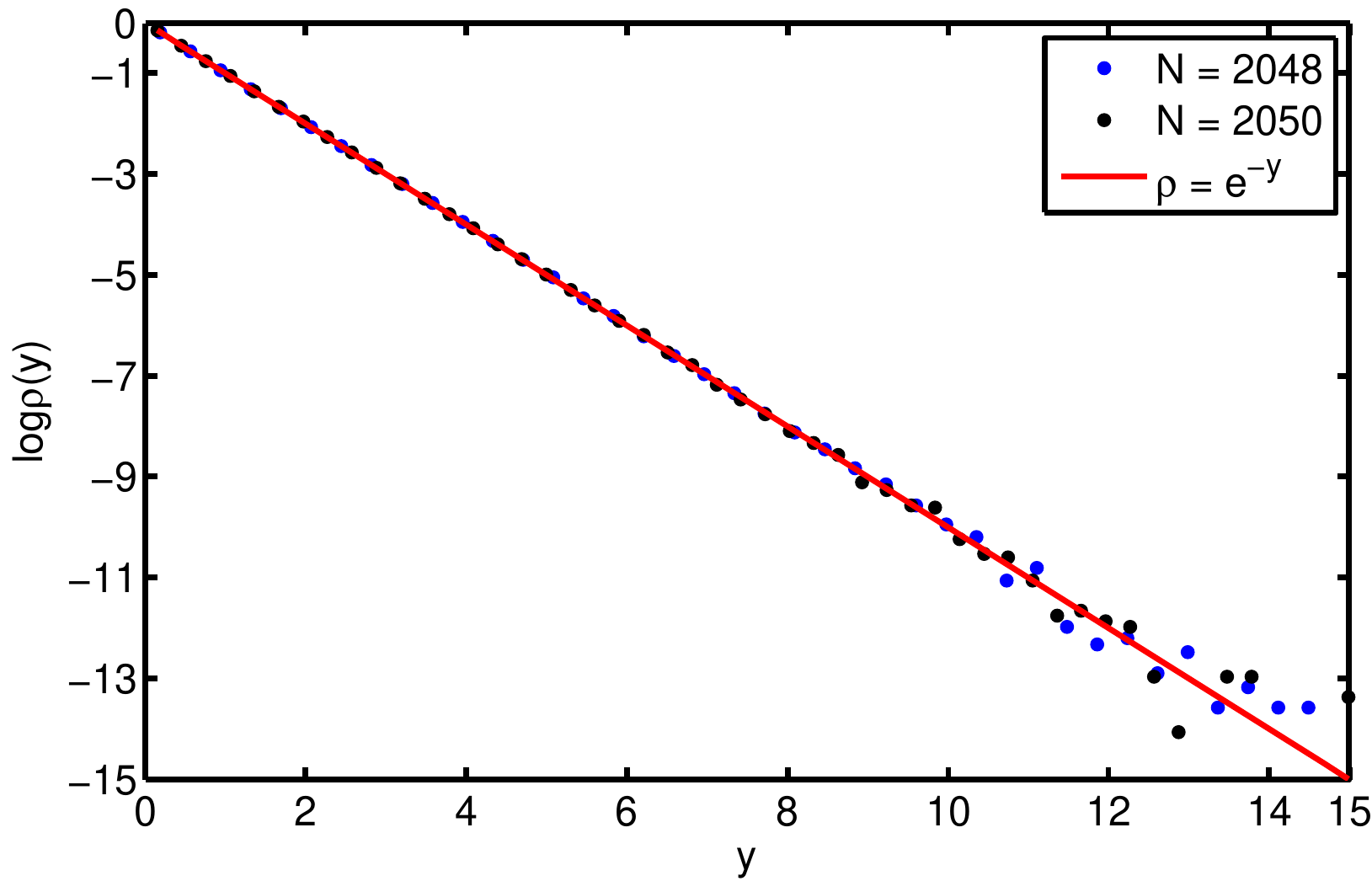}
\includegraphics[scale=0.4]{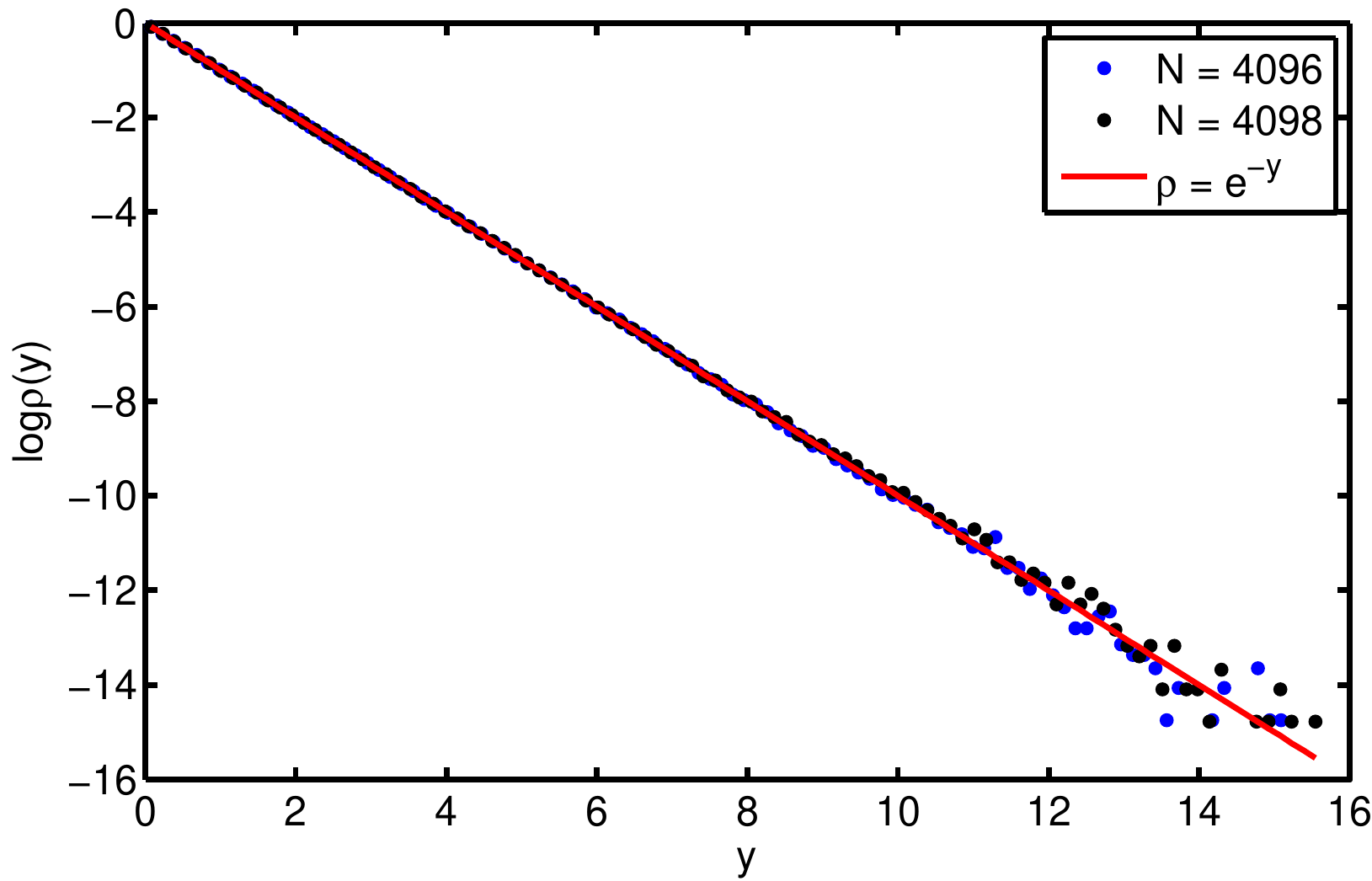}
\end{center}
\caption{Intensity distribution for eigenvectors of standard map at
$K=10$, $\alpha=0.25$, $\beta=0.25$ for different $N=2048,~
2050,~4096,~4098$.}
 \label{fig:chrec:rho_gue}
\end{figure}

Let us briefly recall the intensity distribution expected from random
matrix theory for a complex random state. As the expected value of each
intensity component $x$ in an $N$ dimensional space is $1/N$, it will
be convenient to transform to normalized variable $y=N x$, and in the
limit $N \rightarrow \infty$ its distribution becomes the exponential
distribution,
\beq
\rho(y) = e^{-y}.
\eeq
For $N=2048,~ 2050,~4096,~4098$, this is shown in Fig.
\ref{fig:chrec:rho_gue}. Except the tail region, it matches very well
with the exponential distribution.

It is interesting to study the tail region, and the distribution of large intensities. Is there
a pattern in observing larger intensities (``records") as we go
along the index of eigenvectors? What is the probability of
occurrence of maximum intensity in, say the middle of the eigenfunction?
These are some questions that we will be addressing
comparing with the results obtained for random states in Section
\ref{sec:rec:delseq}. For large $K$ quantum eigenstates of standard map follow the CUE/GUE
or COE/GOE results depending on the value of the phases $\alpha$ and
$\beta$. If $\beta \ne 0$ and $\alpha \ne 0,1/2$ we can expect that
both the time-reversal symmetry and parity symmetry is broken and the
typical eigenstates are expected to be like complex random states.

 The wavefunctions are basis dependent and record statistics will in general depend on the space in which the eigenfunctions are represented. For small
values of $K$ we expect there to be many localized states in the
momentum space while being nearly uniformly distributed in the
position, as the dominant invariant classical phase space structures are rotational KAM tori. However for large $K$, position or momentum basis will be equivalent statistically.
Our analysis below is based on the quantum standard map on the torus.
It is well known that the eigenstates of the quantum standard map on
the cylinder are exponentially localized in momentum space
(\cite{Izr1990}) and share phenomenology similar to Anderson
localization. Study of extremes or records in this case may also be
interesting, but is not pursued further.

\begin{figure}[htb]
\centering
   \begin{subfigure}[t]{0.48\linewidth} \centering

\includegraphics[width=0.98\linewidth]{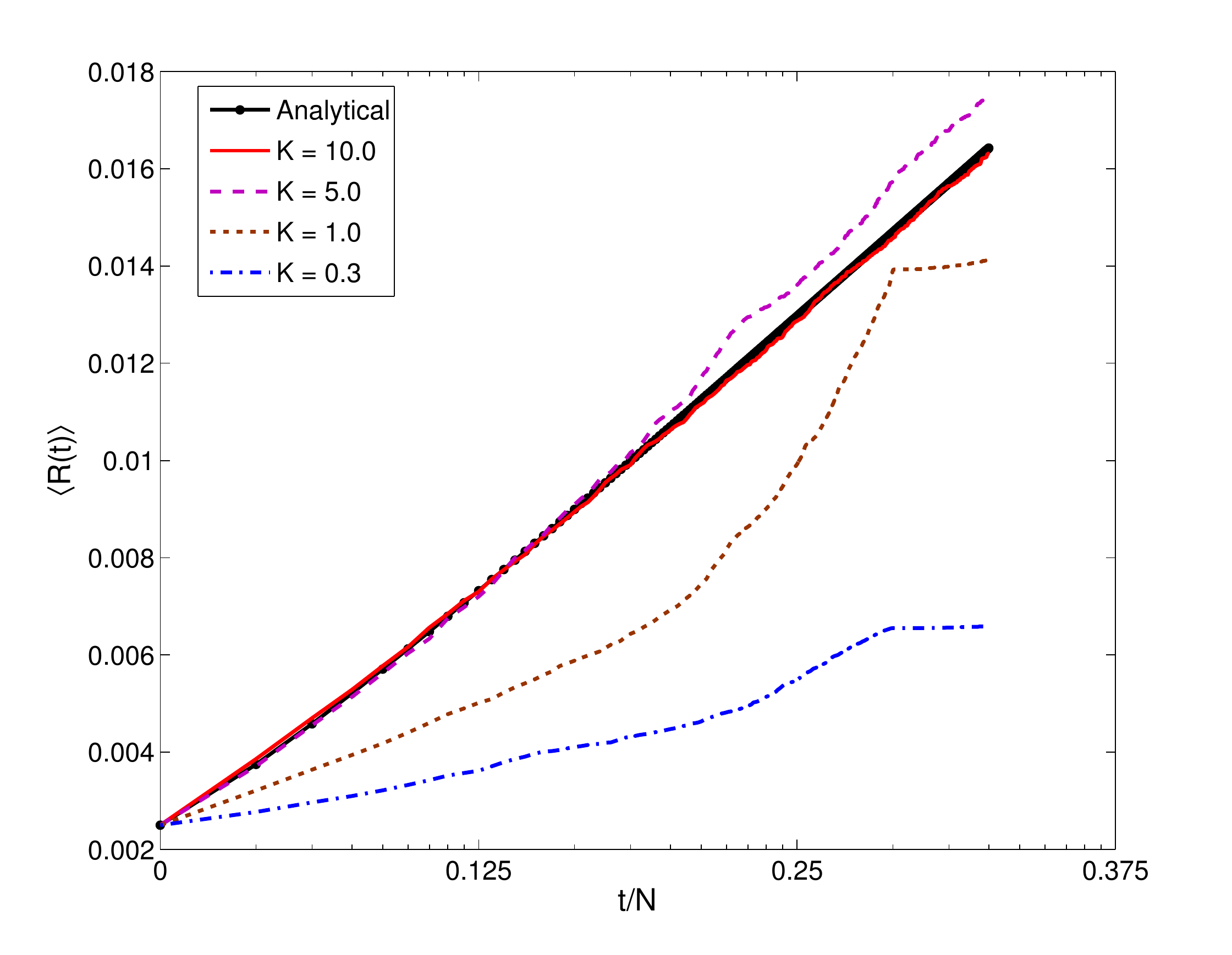}
     \caption{position representation}\label{fig:rt_tbyN}
   \end{subfigure}
   \begin{subfigure}[t]{0.02\linewidth}\centering
   \end{subfigure}
   \begin{subfigure}[t]{0.48\linewidth} \centering

\includegraphics[width=\linewidth]{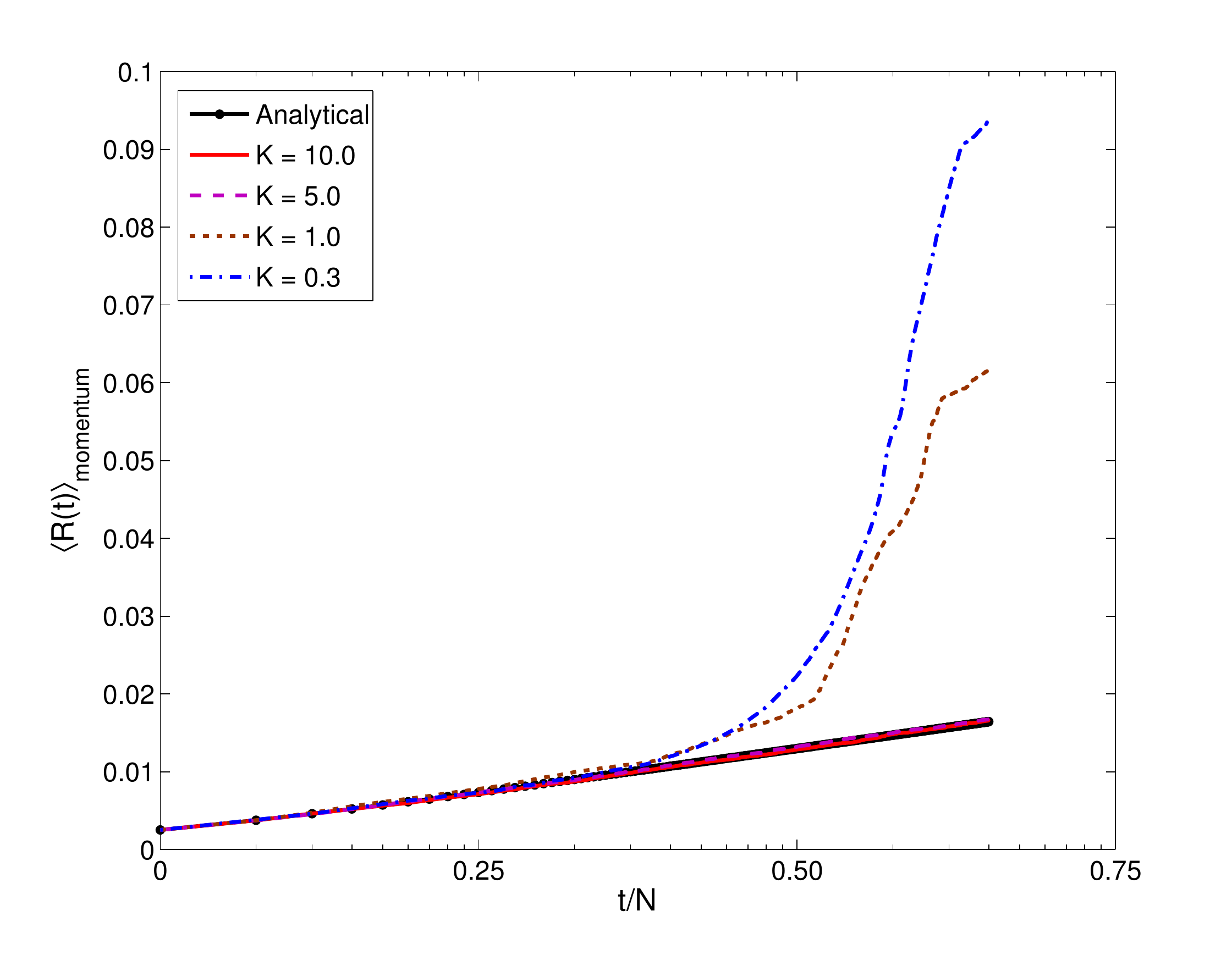}
     \caption{momentum representation}\label{fig:rt_tbyNmom}
   \end{subfigure}
\caption{The average upper record $\langle R(t)\rangle$, from the
ensemble of eigenstates of the quantum standard map. The parameters
used are $N=400$ and $K=10$ (highly chaotic), $K=5$ (mostly chaotic),
$K=1$ (mixed phase space), and $K=0.3$ (mostly regular). The
analytical curve refers to the random state result in Eq.
 (\ref{randomavgrecord}). In all cases $\alpha=\beta=0.25$.}
\label{fig:rt}
\end{figure}

The average value of the upper record as a function of the index for various values of $K$ is plotted in
Fig.~\ref{fig:rt_tbyN}.  This agrees well with the random
states result in the chaotic region for example when $K=10$.
However there are significant deviations in the mixed phase space regime. For example when  $K=0.3$, in the position space most of the records are set up by $t/N=0.5$. It has been pointed out in \cite{slj12arxiv} that this feature is due to the weakly broken parity symmetry. There
are significant deviations from the random state even for $K=5$ when the phase space is largely chaotic. The momentum space
average records  in the mixed phase space regime lie above the random
state result and are not affected so much by the weakly broken parity
symmetry due to their localization (see Fig. \ref{fig:rt_tbyNmom}).
Thus for small $K$ we see much larger records being set than for the
position space.

\begin{figure}[htb]
\begin{center}
\includegraphics[scale=0.6]{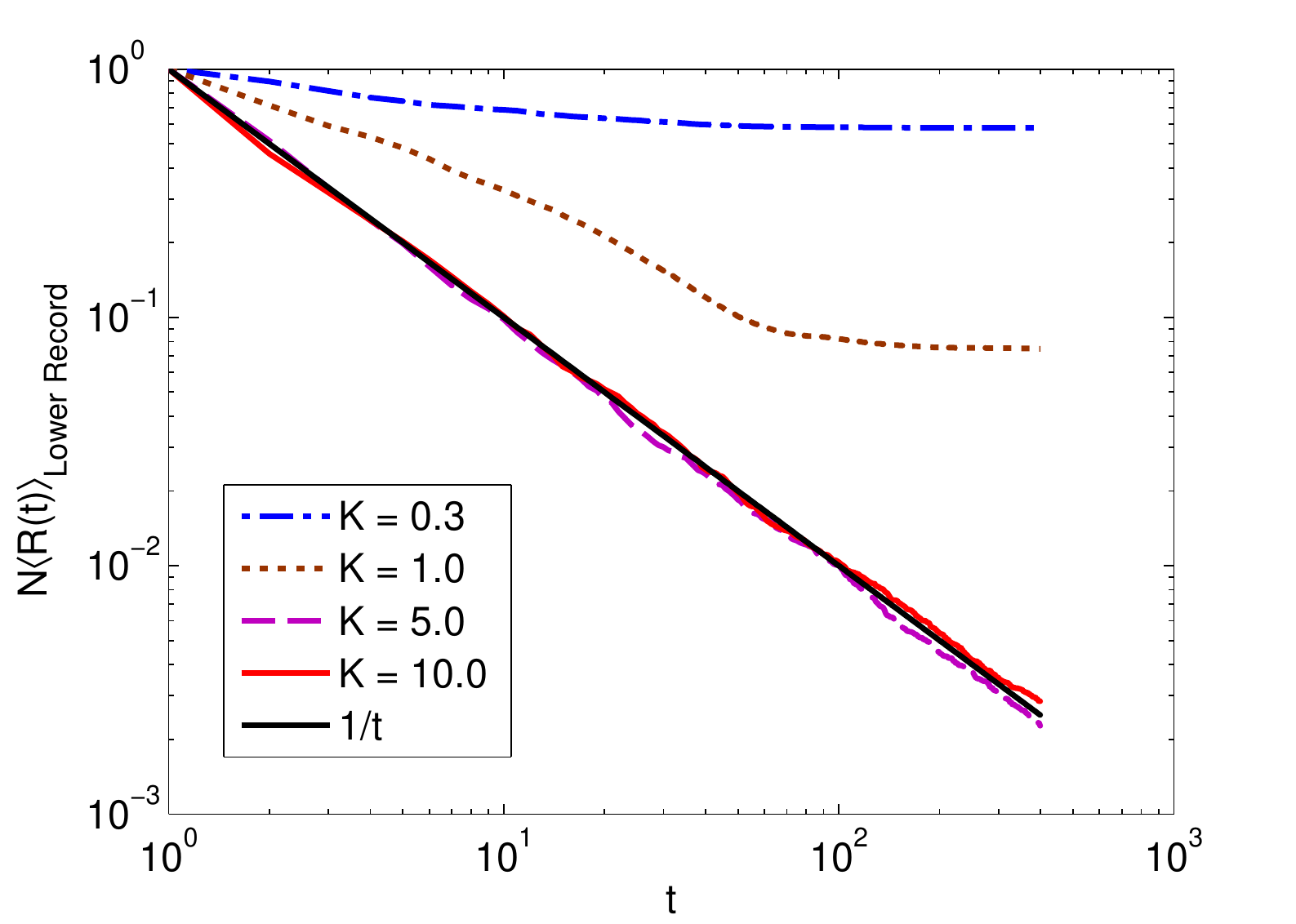}
\end{center}
\caption{The average lower record $\langle R(t)\rangle_{\text{lower
record}}$, from the ensemble of eigenstates of the quantum standard
map in position representation. The parameters used are $N=400$ and
$K=10$ (highly chaotic), $K=5$ (mostly chaotic), $K=1$ (mixed phase
space), and $K=0.3$ (mostly regular). The analytical curve refers to
the random state result in Eq. (\ref{eq:rec:rt_tmin}). In all cases
$\alpha=\beta=0.25$.}
\label{fig:rt_tbyNmin}
\end{figure}

 A similar picture appears with lower records as well, the results of
random vector is followed for large $K(=10)$, while for smaller $K$
the records themselves do not vary with $t$ as compared to chaotic
cases (see Fig. \ref{fig:rt_tbyNmin}). As is clear that upper record
and lower record combined will give the variation of the values that
function assumes, let's call it range, therefore in the case of
standard map for smaller $K$ values (close to integrable regime), say
$K=0.3$, average range of the intensities of eigenvectors is very
small as compared to large $K$ values (chaotic regime), say $K=10$
(See Fig. \ref{fig:rt_tbyN}
and Fig. \ref{fig:rt_tbyNmin}). This is consistent with semiclassical
analysis that  for smaller value of $K$  invariant rotation KAM  tori
are not broken and therefore support wavefunctions with smaller range.

\begin{figure}[htb]
\centering
   \begin{subfigure}[t]{0.48\linewidth} \centering
   \includegraphics[width=0.95\linewidth]{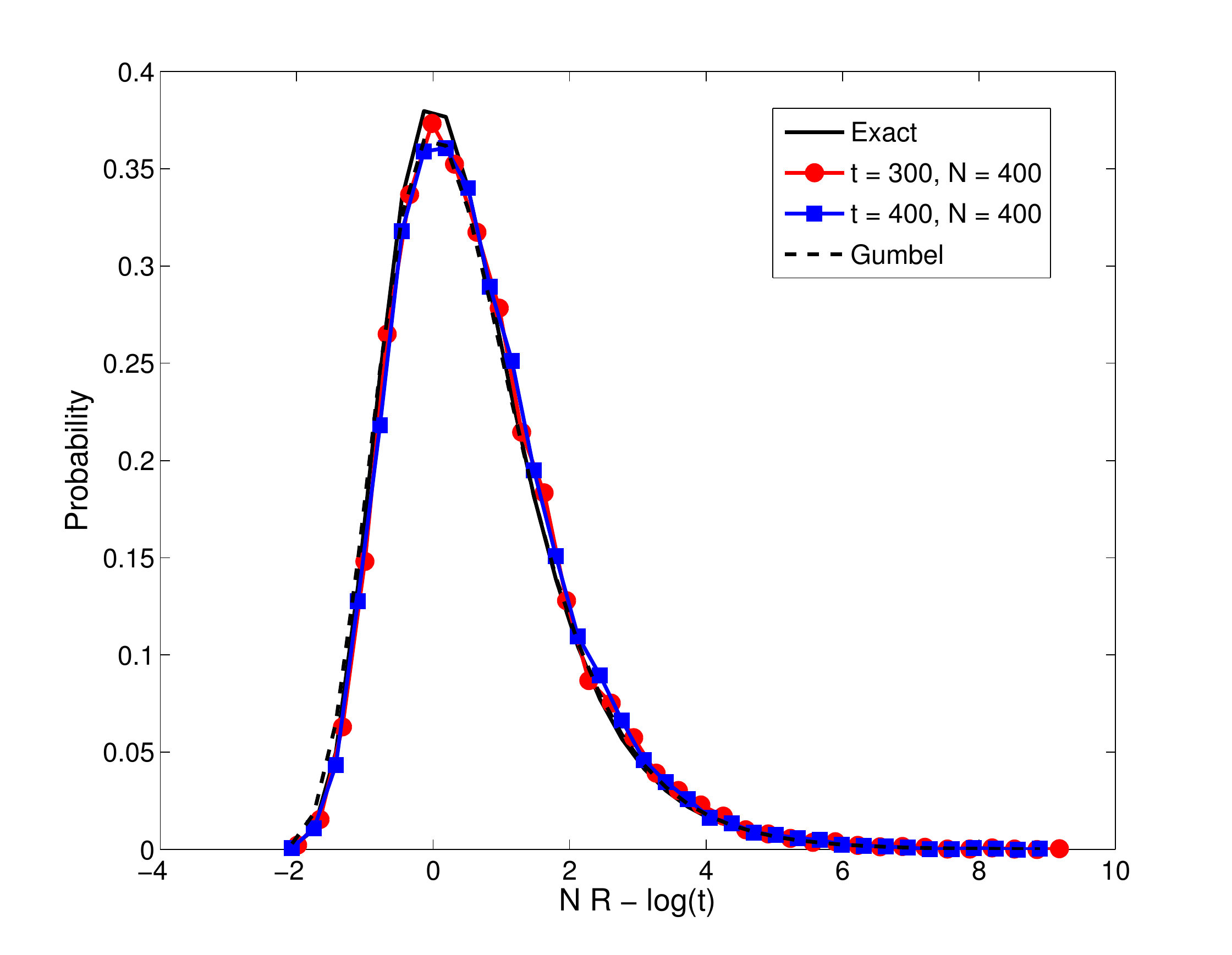}
   \caption{upper record}\label{fig:Gumbel}
   \end{subfigure}
   \begin{subfigure}[t]{0.02\linewidth}\centering
   \end{subfigure}
   \begin{subfigure}[t]{0.48\linewidth} \centering
   \includegraphics[width=\linewidth]{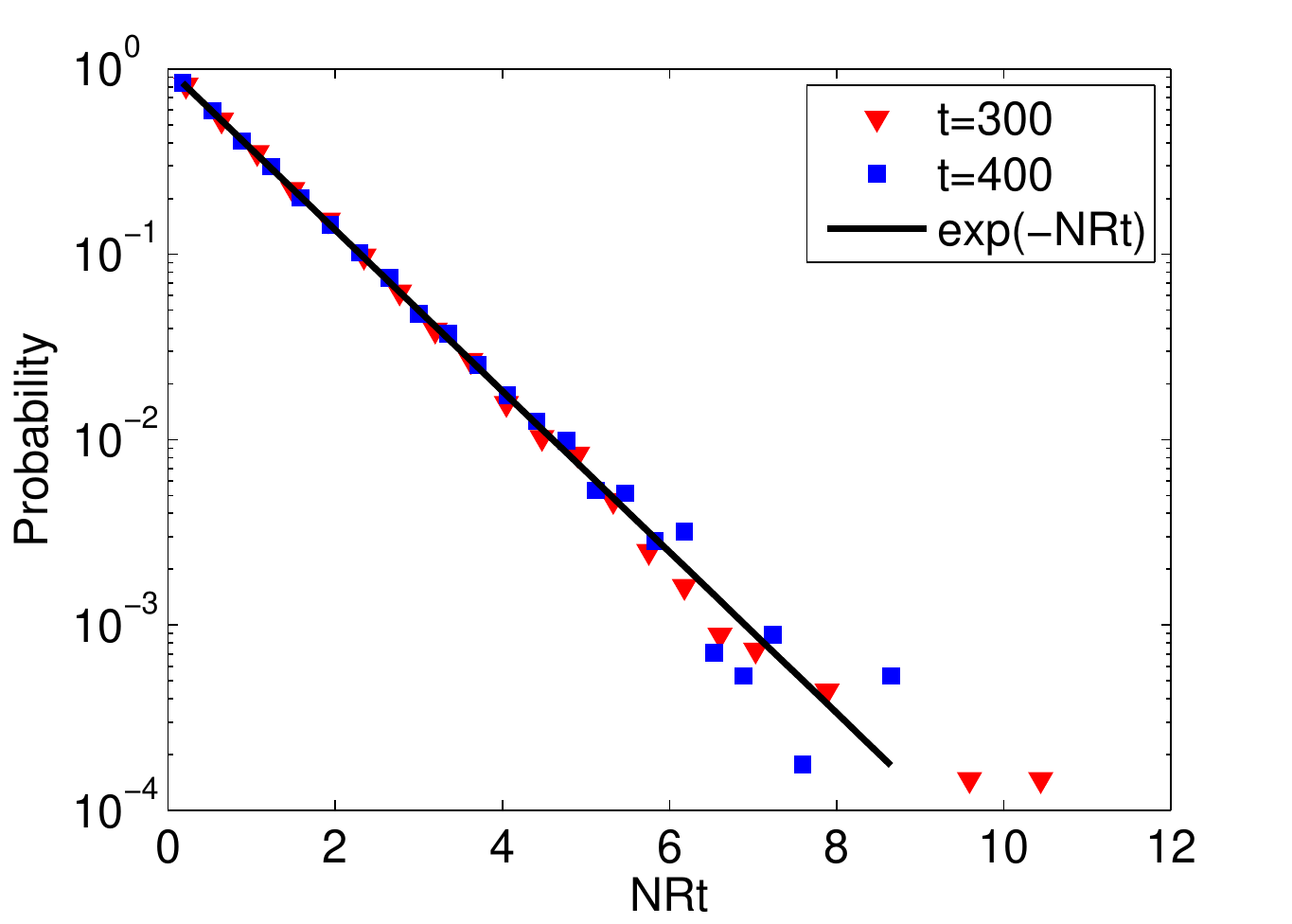}
     \caption{lower record}\label{fig:rec:expo_min}
   \end{subfigure}
\caption{ The distribution of the upper (lower) records when the
index is $t$ for eigenfunctions of the quantum standard map with
$K=10$. After re-scaling and a shift, the distributions are of the
Gumbel (exponential) type. }
\end{figure}

As has been previously discussed, the distribution of the upper
(lower) record at ``time'' $t$ is  Gumbel (exponential) for large $N$
with appropriate shift and scaling.  It is shown in
Fig.~\ref{fig:Gumbel} (Fig. \ref{fig:rec:expo_min}) that indeed the
upper (lower) record for eigenfunctions of the quantum standard map
in the classically chaotic regime is Gumbel
(exponentially)-distributed; also plotted is the distribution for the
``upper (lower) record'' when $t=N$ which refers to the maximum
(minimum) intensity, thus recovering the earlier results of
\cite{ltbm08prl}.

\begin{figure}[htb]
\centering
   \begin{subfigure}[t]{0.48\linewidth} \centering
   \includegraphics[width=0.96\linewidth]{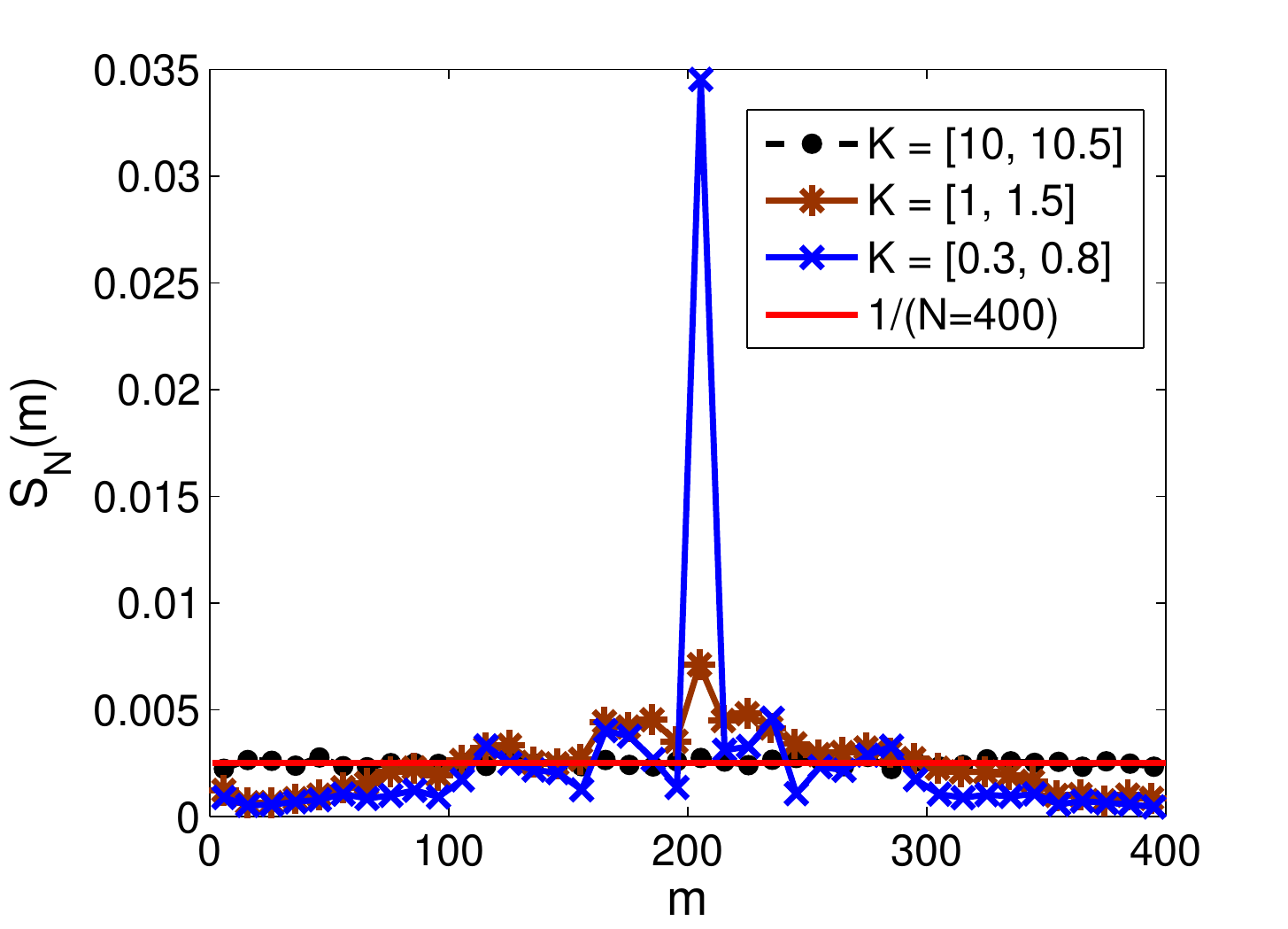}
   \caption{upper record}\label{fig:SN_m}
   \end{subfigure}
   \begin{subfigure}[t]{0.02\linewidth}\centering
   \end{subfigure}
   \begin{subfigure}[t]{0.48\linewidth} \centering
   \includegraphics[width=\linewidth]{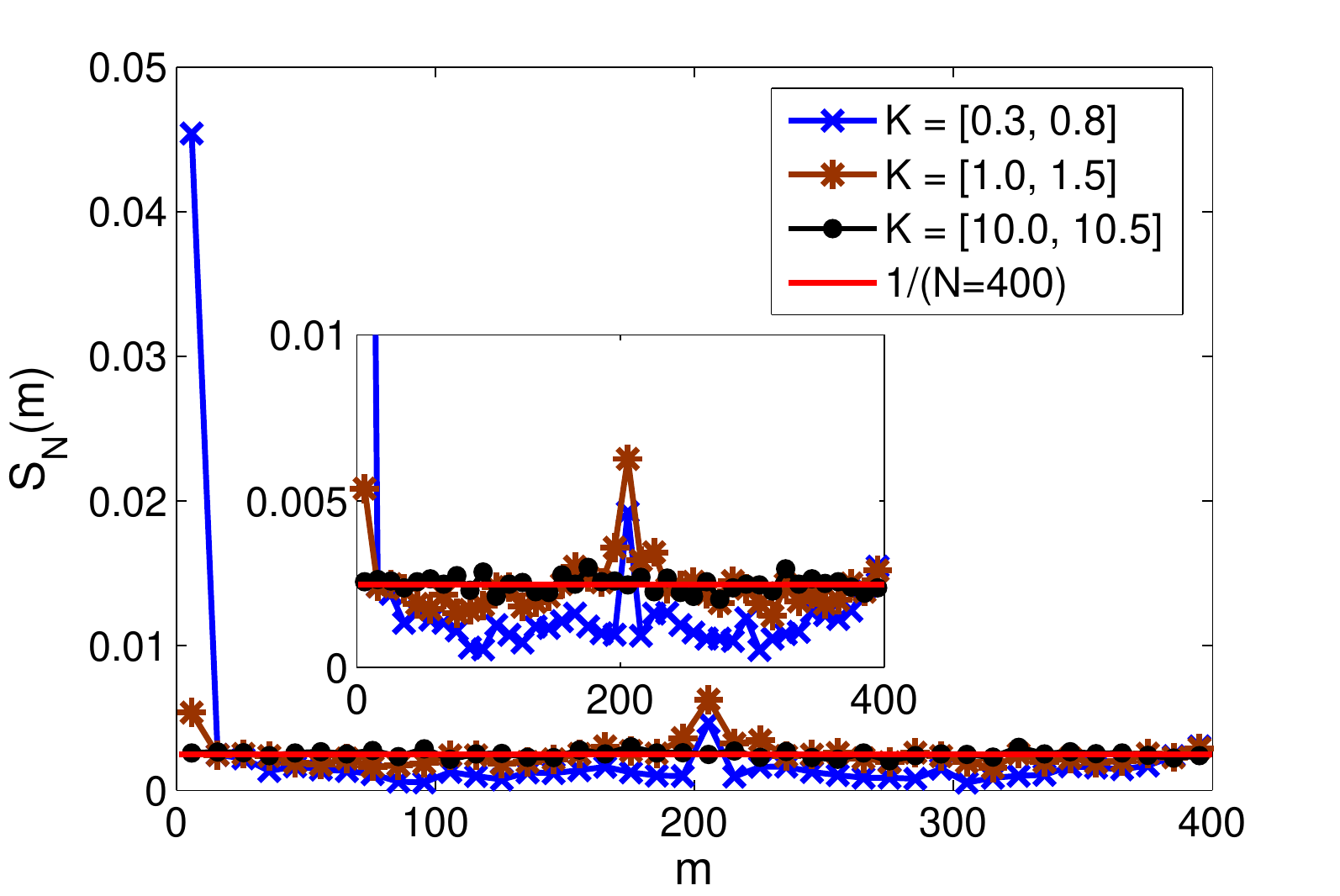}
     \caption{lower record}\label{fig:rec:SNm_min}
   \end{subfigure}
\caption{The distribution of the position of the last record set,
which is also the maximum in case of upper record and minimum in case
of lower record, for eigenfunctions of the standard map with $N=400$
and for various values of $K$. In Fig. \ref{fig:rec:SNm_min}, inset
shows a zoomed in view to clearly see the features of $S_N(m)$ as
function of $m$ for various $K$ values.}
\end{figure}

The distribution of the {\it position} of the maximum intensity in the position
representation is shown in Fig. \ref{fig:SN_m}, where one can see a
transition to the uniform distribution along with the transition to
classical chaos. As has been discussed above, in the context of the
quantity $S_N(m)$, the probability that the final record, which is
also the maximum, lasts for time $m$ is independent of $m$ for both
{\it i.i.d.} and delta correlated variables. Thus the uniform distribution of the maximum intensity is consistent with this.

 The sharp peak at the center for small $K$ (here
$K=0.3$)  deserves further analysis. In \cite{slj12arxiv}, it has been pointed out that for small $K$, when there are many narrow classical resonances, a large fraction of eigenfunctions are localized on separatrices and have
maximum intensity at or very close to $q=1/2$, although this is an unstable fixed point (zero momentum). For example, with $N=400$, $K=0.1$ and $0.3$, about $75\%$ and $50\%$ of eigenstates are
peaked at $q=1/2$.  This is because at the prominent $0/1$ resonance corresponding to the fixed points, $q=1/2$ is near the turning point of orbits on which the eigenstates are localized. 

As $K$ increases there are more prominent resonances and
turning points move away from $q=1/2$. Eigenstates localize in
the resonance interiors, thereby the maximum intensity shifts away from
$q=1/2$. This was the qualitative picture put forward in \cite{slj12arxiv} as leading to the uniform distribution for maximum intensity for large $K$.
 This qualitative explanation is also well supported by the distribution of the
position of the {\it minimum} in the position representation. As we expect and
indeed observe that most of the states have their minimum around
stable fixed point, {\it i.e.} $q=0$ (see Fig. \ref{fig:rec:SNm_min}).
The transition in classical behavior from integrable ($K=0$), to
mixed phase space where islands of stability coexist within the stochastic sea (intermediate $K$), to fully-developed chaos is thus well captured by the quantity $S_N(m)$, the distribution of the position of extreme intensities.

It has been shown above that for random $N$-dimensional states
the number of intensity records increases as in the {\it i.i.d.}
case, namely as  $\sim \log N+\gamma$ both upper and lower records.
While we expect to see this for the quantum standard map
eigenfunctions in the strongly chaotic, large $K$ regime, the mixed
and near-integrable
regimes show a marked departure that were discussed in
\cite{slj12arxiv}  for the case of upper records. The correlations
present for small $K$ lead to results that are similar to those for
the random walks, or the {\it classical} standard map and results in
records increasing as a power-law with
$N$. For example it was found that it is almost a pure power-law with
exponent $0.5$ at the critical value $K = 0.98$, exactly like that of
random walks.

\section{Summary}

We have shown that time evolution of records for squared
deviation of momentum of the classical standard map follows a power law.
In a heuristic manner, it has been argued why in the hard chaos
regime this is the same as the results for discrete random walks with
jump distribution being {\it i.i.d.} random variables.
Hence, any deviation from such a result accounts for the correlation present in dynamical system, and is clearly borne out in the numerical simulations. The deviation for the large $K$ regime is due to the presence of accelerator modes, and result in records that maybe relevant for correlated Levy walks.
Even, in the mixed phase regime the higher period accelerator modes seem to affect the records.

We have reviewed the results on upper records of
intensities of correlated random vectors and derived new ones for lower
records. Apart from deriving the
average record, it has been shown that the probability that a record
appears at an index $j$ is a Bernoulli process, which is the same as
for {\it i.i.d.} variables. The quantum standard map is a good test bed and has increasingly complex spectrum with the
system parameter $K$. For a quantum system with random high-lying
states, records' statistics for random vectors applies.
  Also the study of the position of the last record set in the case of the
standard map, parametrized by $K$ suggests that beyond a certain
value of $K$, the eigenvectors do become like random vectors insofar as
the records of intensities is concerned. This is also consistent
with finding where the number of records vs $N$ goes through a
transition from a power law to logarithmic, as $K$ increases.

We have essentially begun studying classical and quantum dynamical
systems using novel measures related to the theory of extremes and
records, and hope to have shown that this is fruitful and interesting
enough to pursue further.

\section*{Acknowledgement}
We wish to thank Sudhir Jain for the collaboration
that involved  studying the quantum standard map records, and also
gratefully acknowledge Arnd B\"acker for a critical reading of the
manuscript.

\bibliographystyle{elsarticle-harv}
\bibliography{extreme}{}


\end{document}